\documentclass{ieeeaccess}

\usepackage{amsmath,amssymb,amsfonts}
\usepackage{stfloats}
\usepackage{cite}
\usepackage{graphicx}
\usepackage{psfrag}
\usepackage{subfigure}
\usepackage{amsmath}
\usepackage{array}
\usepackage{algorithmicx}
\usepackage{algpseudocode}
\usepackage{setspace}
\usepackage{blindtext}
\usepackage{url}
\usepackage{epstopdf}
\usepackage{verbatim}
\usepackage{color}
\usepackage{amsmath}
\usepackage{bm}
\usepackage{multirow}
\usepackage{booktabs}
\usepackage{makecell}
\usepackage{ntheorem}
\usepackage{textcomp}
\usepackage{caption}
\usepackage{amsmath}

\newtheorem{Thm}{Theorem}

\newtheorem{Prob}{Problem}

\newtheorem{hyp}{Hypothesis}
\newtheorem{prop}{Proposition}

\newcommand{\RNum}[1]{\uppercase\expandafter{\romannumeral #1\relax}}
\newcommand{\tabincell}[2]{\begin{tabular}{@{}#1@{}}#2\end{tabular}}
\theoremseparator{:}

\def\BibTeX{{\rm B\kern-.05em{\sc i\kern-.025em b}\kern-.08em
    T\kern-.1667em\lower.7ex\hbox{E}\kern-.125emX}}

\begin{document}

\history{Received September 1, 2019, accepted September 30, 2019, date of publication October 8, 2019,\\
date of current version October 21, 2019.}
\doi{10.1109/ACCESS.2019.2946271}

\title{Robust Sub-meter Level Indoor Localization with a Single WiFi Access Point - Regression versus Classification}

\author{\uppercase{Chenlu Xiang}\authorrefmark{1}, \IEEEmembership{Student Member, IEEE},
\uppercase{Shunqing Zhang\authorrefmark{1}, \IEEEmembership{Senior Member, IEEE}, Shugong Xu\authorrefmark{1}, \IEEEmembership{Fellow, IEEE}, Xiaojing Chen\authorrefmark{1}, Shan Cao\authorrefmark{1}, \IEEEmembership{Member, IEEE}, George C. Alexandropoulos\authorrefmark{2}, \IEEEmembership{Senior Member, IEEE}, and Vincent Lau}\authorrefmark{3}, \IEEEmembership{Fellow, IEEE}}

\address[1]{Shanghai Institute for Advanced Communication and Data Science, Key laboratory of Specialty Fiber Optics and Optical Access Networks, School of Information and Communication Engineering, Shanghai University, Shanghai, 200444, China (e-mail: \{xcl, shunqing, shugong, jodiechen, cshan\}@shu.edu.cn)}
\address[2]{Department of Informatics and Telecommunications, National and Kapodistrian University of Athens Panepistimiopolis Ilissia, 15784 Athens, Greece, (e-mail: alexandg@di.uoa.gr)}
\address[3]{Department of ECE, Hong Kong University of Science and Technology, Clear Water Bay, Kowloon, Hong Kong S. A. R., China (e-mail: eeknlau@ust.hk)}

\tfootnote{This work was presented in part at the IEEE International Conference on Communications (ICC), Shanghai, China, May 2019 \cite{xiang2019robust}. This work was supported by the National Natural Science Foundation of China (NSFC) Grants under No. 61701293 and No. 61871262, the National Science and Technology Major Project Grants under No. 2018ZX03001009, the Huawei Innovation Research Program (HIRP), and research funds from Shanghai Institute for Advanced Communication and Data Science (SICS). }

\markboth
{C. Xiang \headeretal: Robust Sub-meter Level Indoor Localization with a Single WiFi Access Point - Regression versus Classification}
{C. Xiang \headeretal: Robust Sub-meter Level Indoor Localization with a Single WiFi Access Point - Regression versus Classification}

\corresp{Corresponding author: Shunqing Zhang (e-mail: shunqing@shu.edu.cn).}

\begin{abstract}
Precise indoor localization is an increasingly demanding requirement for various emerging applications, like Virtual/Augmented reality and personalized advertising. Current indoor environments are equipped with pluralities of WiFi access points (APs), whose deployment is expected to be massive in the future enabling highly precise localization approaches. Though the conventional model-based localization schemes have achieved sub-meter level accuracy by fusing multiple channel state information (CSI) observations, the corresponding computational overhead is usually significant, especially in the current multiple-input multiple-output orthogonal frequency division multiplexing (MIMO-OFDM) systems. In order to address this issue, model-free localization techniques using deep learning frameworks have been lately proposed, where mainly classification methods were applied. In this paper, instead of classification based mechanism, we propose a logistic regression based scheme with the deep learning framework, combined with Cram\'{e}r-Rao lower bound (CRLB) assisted robust training, which achieves more robust sub-meter level accuracy (0.97m median distance error) in the standard laboratory environment and maintains reasonable online prediction overhead under the single WiFi AP settings.
\end{abstract}
\titlepgskip=-15pt

\maketitle

\section{Introduction} \label{sect:intro}
Precise indoor localization, a raising demand from our daily lives, brings a brand-new navigation experience \cite{chintalapudi2010indoor} in modern shopping malls or exhibition halls. Since the traditional outdoor positioning systems, such as Global Navigation Satellite Systems (GNSS) \cite{marais2005land}, suffers from the satellite signal blocking effect, some newly deployed infrastructure is often required to achieve high resolution indoor localization accuracy. Typical examples include sound or ultrasonic collection systems \cite{ijaz2013indoor}, Bluetooth Low Energy (BLE) systems \cite{altini2010bluetooth}, radio frequency identification (RFID) receivers and tags \cite{bouet2008rfid}, infrared equipment \cite{hauschildt2010advances}, or even hybrid of them.

Due to the extreme low deployment cost, WiFi access points (APs) are massively deployed in the indoor environment for information transferring, and recently utilized to perform high resolution indoor localization as illustrated in \cite{youssef2005horus,kotaru2015spotfi,vasisht2016decimeter,sen2012you,chapre2015csi}. Compared with aforementioned techniques, the additional deployment cost is usually negligible, and the main challenges nowadays are the high accurate localization algorithms. Among the existing approaches, fingerprint-based schemes have been proven to be an effective solution, where the intrinsic features of WiFi signals are extracted in the training stage and utilized in the operating stage to predict the location through real time measured signals. ``HORUS'', a typical fingerprint-based localization system, has been proposed in \cite{youssef2005horus}, which relies on the received signal strength indication (RSSI) to generate signal features. More accurate localization schemes have been proposed in \cite{kotaru2015spotfi,vasisht2016decimeter,sen2012you,chapre2015csi}, where the real time channel state information (CSI) are measured and processed instead to improve the localization accuracy. For example, SpotFi \cite{kotaru2015spotfi} and Chronos \cite{vasisht2016decimeter} extract the propagation parameters from CSIs, including angle of departure (AOD), angle of arrival (AOA), and time of flight (TOF) information, to compute the relative locations from the reference APs. Another common approach establishes a probabilistic model between the collected CSIs and the candidate locations through some classifiers such as deterministic k-nearest neighbor (KNN) clustering and probabilistic Bayes rule algorithms \cite{sen2012you,chapre2015csi}. The above fingerprint-based solutions are able to achieve sub-meter level accuracy if CSIs from multiple APs \cite{kotaru2015spotfi}, multiple frequency bands \cite{vasisht2016decimeter} or multiple antennas \cite{chapre2015csi} can be fused together. However, the corresponding computational overhead during the offline modeling and online feature extraction is usually significant as shown in \cite{wang2017biloc}.

Apart from the above model-based approaches, the {\em model-free} localization schemes have also been widely investigated during recent years, especially after the deep learning technique has been invented. With the controllable online prediction overhead, the model-free localization approach can directly estimate the corresponding position in the operating/online stage based on the observed and learned relations between the collected CSIs and the labelled locations in the training/offline stage. Typical classification algorithms, including restricted Boltzmann machine (RBM) \cite{wang2017biloc,wang2015deepfi}, convolutional neural networks (CNN) \cite{chen2017confi}, deep residual networks (ResNet) \cite{wang2017resloc}, have been applied to exploit CSI features and classify to different reference positions (RPs) with certain probability. The resultant localization accuracy, after fusing the classification results together, can be significantly improved if compared with the conventional model-based approaches, which ranges from 1.78m to 0.89m in terms of {\em median distance error} (MDE) \cite{wang2017biloc,wang2015deepfi,chen2017confi,wang2017resloc}. The model-free localization scheme partially solves the computational complexity issue, while the localization accuracy is still insufficient for many indoor applications especially when the infrastructure is insufficient. In this paper, we consider a standard laboratory environment with single WiFi AP settings and propose to use a logistic regression based solution \cite{peduzzi1996simulation} instead of using the commonly adopted classification based scheme. Since the regression based scheme can directly model the continuous localization function, it is capable of achieving sub-meter level accuracy (0.97m MDE) in 8m $\times$ 6m room space. In addition, based on the proposed framework, we derive the lower bound of localization errors using Cram\'{e}r-Rao lower bound (CRLB) \cite{kay1993fundamentals} and figure out that a small perturbation in the training stage can eventually help us to reduce the localization errors. We hope the proposed logistic regression based framework can shed some light on the model-free as well as the model-based localization techniques and pave the way for the deep learning based localization algorithms in the practical WiFi MIMO-OFDM systems. The main contributions of this paper are listed below.
\begin{itemize}
\item{\bf Regression versus Classification.}
A straightforward idea to solve the localization problems using deep learning is to extract the features in the operating stage and compare with the pre-collected features of RPs in the training stage. A classification process is then applied to calculate the similarity with respect to different pre-defined RPs. This approach greatly reduces the computational resources for online feature comparison, while the corresponding localization accuracy will be affected due to the limited training space offered by the finite number of RPs. To achieve a better trade-off between the localization accuracy and the online inference capability, a reasonable approach is to expand the finite location set of RPs to the continuous set of the entire room space, where a logistic regression method can be applied.
\item{\bf Unified Optimization Framework.} To provide a detailed understanding of the proposed scheme, we establish a general mapping relationship between the real time measured CSI and the corresponding locations according to the parametric system model. On this basis, we introduce a unified optimization framework to formulate the localization problems using WiFi fingerprints, including both classification and regression based formulations. Based on that, we explain why the logistic regression based approach achieves better localization accuracy than the traditional classification based approaches, and discuss the potential impacts with different system configurations.
\item{\bf CRLB Assisted Robust Training.} Based on the proposed framework, we conduct extensive CRLB analysis to obtain an in-depth understanding of the localization errors in the proposed system. In addition, we show through CRLB analysis that a small perturbation in the training stage can help to accommodate the randomness induced by temporal spatial variation, which eventually improves the robustness of the proposed scheme. Therefore, a more robust training strategy is to construct the training dataset for each RP using collected CSIs from this RP and its neighboring areas. As we show through extensive numerical experiments, the proposed CRLB assisted robust training method is able to improve the localization accuracy about 30\%, if compared with the conventional training strategies.
\end{itemize}

The rest of paper is organized as follows. In Section~\ref{sect:pre}, we provide some background information regarding the channel model and the mapping functions. The regression based localization formulation is discussed in Section~\ref{sect:prob} and the corresponding CRLB derivation is provided in Section~\ref{sect:CRLB}. We propose the classification and logistic regression based solutions in Section~\ref{sect:sys} and present our experimental results in Section~\ref{sect:experiment}. Finally, we conclude this paper in Section~\ref{sect:conc}.

\section{Preliminaries} \label{sect:pre}
In this section, we introduce a multipath MIMO-OFDM channel model for the indoor localization environment and then establish the relationship between the channel fingerprints and location information.
\subsection{Channel Model}
\begin{figure}
\centering
\includegraphics[width = 3.5 in]{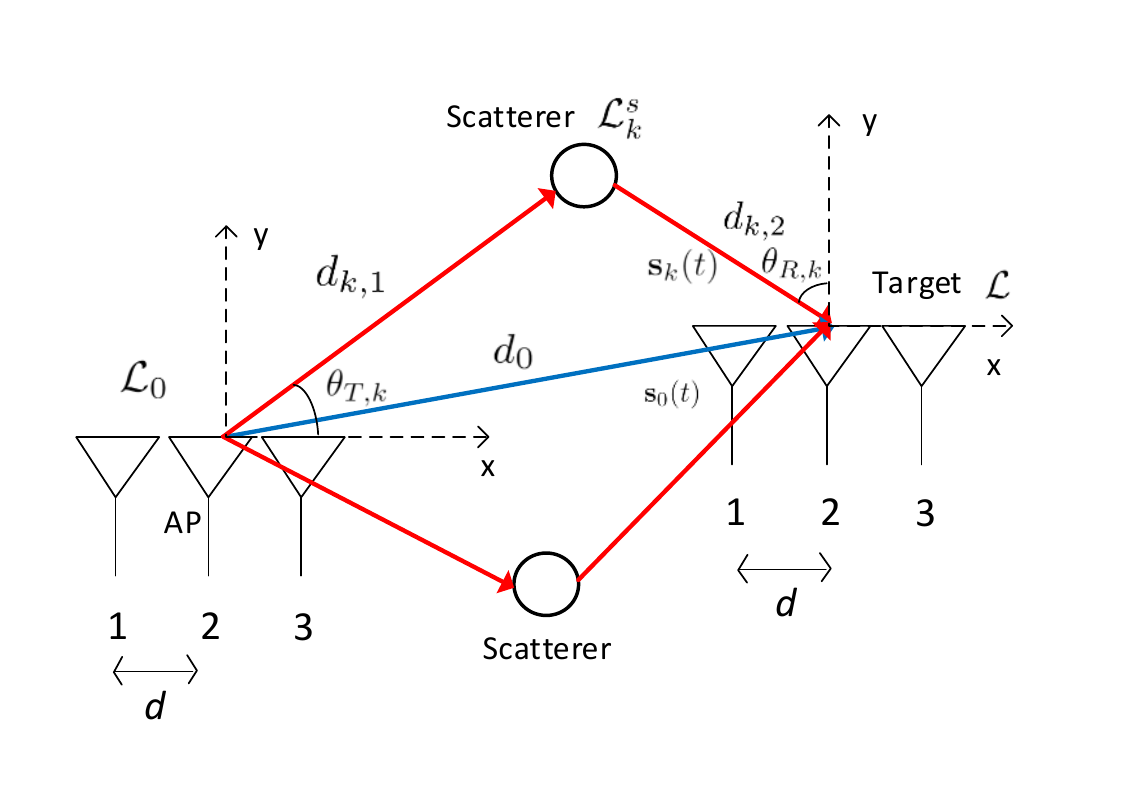}
\caption{Two dimensional illustration of the position-related parameters of the channel. The LOS path and NLOS paths are marked in blue and red respectively.}
\label{fig:multieffect}
\end{figure}

Consider a MIMO-OFDM system with $N_{T} \times N_{R}$ antenna configuration\footnote{The proposed approach is equally applicable to single antenna users by extending the received signals to $N_R$ copies.} as shown in Fig.~\ref{fig:multieffect}, where $N_T$ and $N_R$ represents the number of the transmitted antennas and received antennas. The received signal at the $i^{th}$ subcarrier and the $n^{th}$ OFDM symbol can be modeled through,
\begin{eqnarray}
\mathbf{y}_{i}(\mathcal{L}, n) = \mathbf{H}_{i}(\mathcal{L}, n) \mathbf{x}_{i}(\mathcal{L}, n) + \mathbf{n}_{i}(\mathcal{L}, n),
\end{eqnarray}
where $\mathcal{L}$ is the target location, $\mathbf{y}_{i}(\mathcal{L}, n) \in \mathbb{C}^{N_R}, \mathbf{x}_{i}(\mathcal{L}, n) \in \mathbb{C}^{N_T} $ denote the received and transmitted signal, and $\mathbf{n}_{i}(\mathcal{L}, n)\in \mathbb{C}^{N_R}$ denotes the additive white Gaussian noise, respectively. $\mathbf{H}_{i}(\mathcal{L}, n) \in \mathbb{C}^{N_R \times N_T}$ denotes the collected the corresponding CSI and the overall aggregated channel response $\mathbf{H}(\mathcal{L}, n) \in \mathbb{C}^{N_T \times N_R \times N_{sc}}$ is given by,
\begin{eqnarray}
\mathbf{H}(\mathcal{L}, n) = \Big[
\mathbf{H}_{1}(\mathcal{L},n) \
\mathbf{H}_{2}(\mathcal{L},n) \
\cdots \
\mathbf{H}_{N_{sc}}(\mathcal{L},n) \Big].
\end{eqnarray}

For illustration purpose, we assume uniform linear arrays (ULA) with inter-antenna spacing $d$ are equipped at the transmitter and receiver sides, and all the antenna arrays share the same plane as illustrated in Fig.~\ref{fig:multieffect}. In addition, we consider $K+1$ multiple fading paths in this environment, and the channel responses $\mathbf{H}_{i}(\mathcal{L},n)$ is assumed to remain constant during the transmission of the $n_{th}$ OFDM symbol, which is given by \cite{shahmansoori2017position},
\begin{eqnarray}
\mathbf{H}_{i}(\mathcal{L},n) = \mathbf{A}_{R,i}(\mathcal{L},n)\mathbf{\Gamma}_{i}(\mathcal{L},n)\mathbf{A}^{\textrm{H}}_{T,i}(\mathcal{L},n),
\end{eqnarray}
where the angular domain correlation matrices, $\mathbf{A}_{R/T,i}(\mathcal{L})\in \mathbb{C}^{N_{R/T}\times (K+1)}$, are defined as,
\begin{eqnarray}
\mathbf{A}_{R/T,i}(\mathcal{L}) = [\mathbf{a}_{R/T,i}(\theta_{R/T,0}), \cdots, \mathbf{a}_{R/T,i}(\theta_{R/T,k}), \nonumber\\
\cdots, \mathbf{a}_{R/T,i}(\theta_{R/T,K})],
\end{eqnarray}
with $\mathbf{a}_{R/T,i}(\theta_{R/T,k})= \frac{1}{\sqrt{N_{R/T}}}
\Big[e^{-j\frac{N_{R/T}-1}{2}\frac{2\pi}{\lambda_{i}}d\sin(\theta_{R/T,k})}$, $\cdots, e^{j\frac{N_{R/T}-1}{2}\frac{2\pi}{\lambda_{i}}d\sin(\theta_{R/T,k})}\Big]^{\textrm{T}}$. As shown in Fig.~\ref{fig:multieffect}, we define $\theta_{T,k}$, $\theta_{R,k}$, $d_{k}$ and $\mathcal{L}^{s}_{k}$ as the AOD, AOA, path length and the unknown scatterer location of the $k^{th}$ path\footnote{When $k = 0$ for the LOS path, $d_k= c\cdot\tau_k$, where $\tau_k$ is TOF and $c$ is the speed of light.}. Then we have $d_{k}^{1} = ||\mathcal{L}^{s}_{k} - \mathcal{L}_{0}||_2$ and $d_{k}^{2} = ||\mathcal{L} - \mathcal{L}^{s}_{k}||_2$, where $\mathcal{L}_{0}$
is the known AP location. Additionally, $\mathbf{\Gamma_{i}}(\mathcal{L}) = \sqrt{N_{T}N_{R}} \times \textrm{diag}\left\{h_{0}e^{\frac{-j2\pi i \tau_{0}}{N_{sc}T_{s}}}, \cdots, h_{k}e^{\frac{-j2\pi i\tau_{k}}{N_{sc}T_{s}}}, \cdots, h_{K}e^{\frac{-j2\pi i\tau_{K}}{N_{sc}T_{s}}}\right\}$, where $h_{k}$ is the channel coefficient of the $k^{th}$ path and $T_s$ is the sampling period.

\subsection{Function Mapping Relationship}
Due to the multipath effect, the channel matrix $\mathbf{H}(\mathcal{L}, n)$ is also affected by the indoor environment (objects placement or humans movement), except for the location $\mathcal{L}$. That is a certain location $\mathcal{L}$ corresponds to a channel state matrix set $\{\mathbf{H}(\mathcal{L}, n)\}$ as its fingerprints, which is kind of one-to-many function mapping relationship. We denote the relationship between fingerprints $\mathbf{H}(\mathcal{L}, n)$ and the location $\mathcal{L}$ as function $f(\cdot)$, which is expressed as,
\begin{eqnarray}
\mathbf{H}(\mathcal{L}, n) = f\left(\mathcal{L}\right), \forall n \in [1,N],
\end{eqnarray}
where $N$ denotes the total number of OFDM symbols in each localization positions.
It is a natural question to ask if $\mathbf{H}(\mathcal{L}, n)$ can be mapped to a certain location $\mathcal{L}$ and what is the condition of the inverse function exists.
\begin{hyp} \label{hyp:first}
When the perfect channel knowledge $\mathbf{H}(\mathcal{L}, n)$ is obtained, the observation CSI can be mapped to a certain position.
\end{hyp}
\begin{IEEEproof}[Proof of Hypothesis~\ref{hyp:first}]
Please refer to Appendix~\ref{appendix:proof_hyp} for the proof.
\end{IEEEproof}

In the conventional model-based approach, we are supposed to find the closed-form relationship between $\mathbf{H}(\mathcal{L}, n)$ and $\mathcal{L}$ by exploiting AOA and TOF features, which is generally complicated. While in the model-free based scheme, we directly figure out the characteristics of the localization function $g(\cdot)$ and propose a better approximation by learning the collected CSIs and locations in the training stage.

We make the following assumptions in the rest of the paper. Firstly, we assume availability of perfect CSI measurements leaving extensions to imperfect CSI, due to e.g. imperfect hardware components and limited pilot power, as well as low noise cases for future work. Secondly, we only collect the CSIs from some discrete RPs, instead of sampling the entire fading environments, to control the deployment complexity. Thirdly, we assume that the re-reflected signal (secondary reflection) is too weak to be considered in the channel model. Last but not least, since the mathematical representation of {\em median distance error} is in general complicated, we define {\em mean distance error} (MDE) \cite{bevington1993data} instead as the performance measurement in the training stage, as well as the loss function design in the training stage.

\section{Problem Formulation}\label{sect:prob}
In this section, we apply a general optimization framework to describe the localization problem. Denote $\mathcal{L}_m$ and $\hat{\mathcal{L}}_m$ to be the ground-true and the predicted locations of the $m^{th}$ target respectively, and the corresponding MDE performance over $M$ sampling positions is given by $\frac{1}{M} \sum_{m=1}^{M} \|\hat{\mathcal{L}}_m - \mathcal{L}_m\|_2$, where $\|\cdot\|_2$ represents the vector $l_2$ norm as defined in \cite{peduzzi1996simulation}, i.e., $\forall x = [x_1, x_2, \ldots, x_K ] \in \mathbb{R}^{K}, \|x\|_2=(\sum_{k=1}^{K}|x_k|^2)^{\frac{1}{2}}$. For illustration convenience, we denote the inverse function of $f(\cdot)$ to be $g(\cdot)$, i.e., $g(\cdot)=f^{-1}(\cdot)$, and the mathematical expression for the location estimation is defined by,
\begin{eqnarray}
\mathcal{L} = g\left(\{\mathbf{H}(\mathcal{L}, n), \forall n \in [1,N] \}\right).
\end{eqnarray}
With the above notation, we can describe the MDE minimization problem using the following optimization framework.
\begin{Prob}[MDE Minimization]
\label{prob:MDE_min}
\begin{eqnarray}
\underset{g (\cdot)}{\textrm{minimize}} && \frac{1}{M} \sum_{m=1}^{M} \|\hat{\mathcal{L}}_m - \mathcal{L}_m\|_2 \\
\textrm{subject to} && \hat{\mathcal{L}}_{m} = g \left(\left\{\mathbf{H}(\mathcal{L}_{m},n_{m})\right\}\right), \\
&& \hat{\mathcal{L}}_{m}, \mathcal{L}_{m} \in \mathcal{A}, \forall m, \label{eqn:const_1}
\end{eqnarray}
\end{Prob}
where $\mathcal{A}$ represents the feasible indoor localization areas and $n_m \in [1, N]$ denotes the duration of the $m^{th}$ localization period with $N$ observed OFDM symbols.

Since the above minimization needs to be evaluated over all the possible choices of functions $g(\cdot)$, conventional classification based approaches decompose the original problem into two stages, where it computes the likelihood functions with respect to several RPs in the first stage and simply applies some basic fusion techniques to obtain the final results in the second stage. The corresponding mathematical formulation is given below.

\begin{Prob}[Classification based Localization]
\label{prob:CBL}
\begin{eqnarray}
\underset{g_1 (\cdot), g_2 (\cdot)}{\textrm{minimize}} && \frac{1}{M} \sum_{m=1}^{M} \|\hat{\mathcal{L}}_m - \mathcal{L}_m\|_2 \\
\textrm{subject to} && \hat{\mathcal{L}}_{m} = g_1 \left(\left\{\hat{\mathcal{L}}_{m} (n_m) \right\}\right),\\
&& \hat{\mathcal{L}}_{m} (n_m) = \mathbf{p}_m^{T} (n_m) \cdot \overline{\mathcal{L}_{RP}}, \\
&& \mathbf{p}_m (n_m) = g_2 \left(\mathbf{H}(\mathcal{L}_{m},n_{m}), \overline{\mathcal{L}_{RP}} \right), \\
&& \mathbf{p}_m (n_m) \in [0,1]^{N_{RP}},\\
&& \mathcal{L}_{m} \in \mathcal{A}, \forall m,
\end{eqnarray}
where $\overline{\mathcal{L}_{RP}}$ and $\mathbf{p}_m (n_m)$ denote the collected locations of all the possible RPs and the likelihood distribution with respect to $\overline{\mathcal{L}_{RP}}$ based on the $n_m^{th}$ OFDM symbol, respectively. $N_{RP}$ represents the number of RPs during the localization process\footnote{In the practical deployment, we choose $M$ to be equal to $N_{RP}$ in order to reduce the testing and data processing complexity for modeling $g_2(\cdot)$ in the training stage.}.
\end{Prob}

In the formulation of Problem~\ref{prob:CBL}, $g(\cdot)$ has been decomposed into two simplified functions, $g_1(\cdot)$ and $g_2(\cdot)$, and the existing literature focuses on modeling $g_2(\cdot)$ as a typical classification problem. $g_1(\cdot)$ usually adopts the mathematical average operation or some Kalman filtering \cite{cai2017cril} based techniques to fuse multiple classification results together. Through this approach, the searching space of candidate location set as well as the corresponding computational complexity can be greatly reduced, e.g. from all feasible location area $\mathcal{A}$ as defined in Problem~\ref{prob:MDE_min} to $\overline{\mathcal{L}_{RP}}$ with finite dimension, $N_{RP}$. However, the above decomposition approach sacrifices the localization accuracy by enforcing the candidate location set to be finite RPs and their trivial combinations. A more reasonable approach is to directly model the function $g^{\star}(\cdot)$ using the logistic regression concept \cite{peduzzi1996simulation}, where we formulate Problem~\ref{prob:RBL} to approximate the original non-convex function $g^{\star}(\cdot)$ for MDE minimization using logistic regression.

\begin{Prob}[Regression based Localization]
\label{prob:RBL}
\begin{eqnarray*}
g^{\star}_{LR}(\cdot) \approx g^{\star}(\cdot) = & \arg \min_{g(\cdot)} & \frac{1}{M} \sum_{m=1}^{M} \|\hat{\mathcal{L}}_m - \mathcal{L}_m\|_2 \\
& \textrm{subject to} & \hat{\mathcal{L}}_{m} = g \left(\left\{\mathbf{H}(\mathcal{L}_{m},n_{m}) \right\}\right), \\
&& \hat{\mathcal{L}}_{m}, \mathcal{L}_{m} \in \mathcal{A}, \forall m,
\end{eqnarray*}
where $g^{\star}_{LR}(\cdot)$ denotes the associated regression function.
\end{Prob}

Since we can search over a larger optimization space of the function $g(\cdot)$, the logistic regression based scheme shall be able to achieve better localization accuracy than the classification based scheme. To control the potential processing complexity for evaluating different functions of $g(\cdot)$, traditional schemes usually rely on the Gaussian regression, which fit the approximation function by calculating means and variances. However, the Gaussian regression approach has the robustness issue and a model-free deep learning based localization approach is more preferable as elaborated in \cite{belagiannis2015robust}. To control the potential deployment complexity associated, we further tighten the constraint \eqref{eqn:const_1} and have,
\begin{eqnarray*}
g^{\star}_{LR}(\cdot) = & \arg \min_{g(\cdot)} & \frac{1}{M} \sum_{m=1}^{M} \|\hat{\mathcal{L}}_m - \mathcal{L}_m\|_2 \\
& \textrm{subject to} & \hat{\mathcal{L}}_{m} = g \left(\left\{\mathbf{H}(\mathcal{L}_{m},n_{m}) \right\}\right), \\
&& \hat{\mathcal{L}}_{m}, \mathcal{L}_{m} \in \overline{\mathcal{L}_{RP}}, \forall m.
\end{eqnarray*}
Kindly note that the above approximation can be improved when the number of RPs, $N_{RP}$, increases, which actually provides a meaningful trade-off between the implementation complexity and the localization accuracy\footnote{When the number of RPs, $N_{RP}$, tends to infinity, we are able to characterize the function $g^{\star}(\cdot)$ in target area $\mathcal{A}$ with probability 1 by learning the function $g^{\star}_{LR}(\cdot)$. However, in the practical systems, we observe that the localization accuracy saturates when $N_{RP}$ exceeds some threshold value.}.

\section{CRLB For localization error}\label{sect:CRLB}
In this section, we present the CRLB of the formulated problem in Section~\ref{sect:prob}, which is widely used to derive a lower bound on the variance of unbiased estimators \cite{amemiya1985advanced}. Fisher Information Matrix (FIM) as defined in \cite{frieden2004science} is utilized to evaluate the CRLB of localization error and the effect of perturbation is analyzed in what follows.

\subsection{CRLB of Localization Error}
In the above formulation, the position of the $m^{th}$ location corresponds to a 2-D location coordinate\footnote{We focus on 2-D position case in this paper, $\mathcal{L}_{m}=(x_m,y_m,z_m) \in \mathbb{R}^{3}$ in case of 3-D position.}, e.g., $\mathcal{L}_{m}=(x_m, y_m) \in \mathbb{R}^{2}$. Denote $\bm{\eta} = \left[\bm{\eta}_{0}^{T},\bm{\eta}_{1}^{T},\cdots\bm{\eta}_{k}^{T},\cdots\bm{\eta}_{K}^{T}\right]^{T}$ to be the collections of unknown parameters for $K+1$ channel fading paths. For the $k^{th}$ fading path, $\bm{\eta}_{k}^{T}\in \mathbb{R}^{N_{\bm{\eta}}}$ represents $N_{\bm{\eta}}$ unknown fading parameters, including delay, angles and channel coefficients. Mathematically, it can be expressed as
\begin{eqnarray}
\bm{\eta}_{k}^{T}=\left[\tau_{k},\bm{\theta}_{k}^{T},\mathbf{h}_{k}^{T}\right]^{T},
\end{eqnarray}
where $\bm{\theta}_{k}=[\theta_{T,k}, \theta_{R,k}]^{T}$ is the collections of AoD and AoA, and $\mathbf{h}_{k} = [\mathbf{h}_{R,k}, \mathbf{h}_{I,k}]^{T}$ contains the real and imaginary parts of the channel coefficients. If we define $\hat{\bm{\eta}}$ as the unbiased estimator of $\bm{\eta}$, the associated covariance matrix, $\textrm{Cov}(\bm{\eta})$, is given by,
\begin{eqnarray}
\textrm{Cov}(\bm{\eta}) = \mathbb{E}_{\mathbf{y},\bm{\eta}}\left[ (\hat{\bm{\eta}}-\bm{\eta})(\hat{\bm{\eta}}-\bm{\eta})^{T}\right]\geq \mathbf{J}_{\bm{\eta}}^{-1},
\end{eqnarray}
where $\mathbf{J}_{\bm{\eta}}$ is the FIM for $\bm{\eta}$. According to the definition of FIM \cite{frieden2004science}, $\mathbf{J}_{\bm{\eta}}$ can be computed via,
\begin{eqnarray}
\mathbf{J}_{\bm{\eta}} &\triangleq& -\mathbb{E}_{\mathbf{y},\bm{\eta}}\left[\frac{\partial^2 \ln p(\mathbf{y}|\boldsymbol{\bm{\eta}})}{\partial\bm{\eta}\partial\bm{\eta}^{\textrm{T}}}\right] \nonumber\\
& = &\left[
\begin{array}{c c c}
\psi(\bm{\eta}_{0},\bm{\eta}_{0})&\cdots &\psi(\bm{\eta}_{0},\bm{\eta}_{K})\\
 \vdots  & \ddots  & \vdots  \\
\psi(\bm{\eta}_{K},\bm{\eta}_{0})&\cdots &\psi(\bm{\eta}_{K},\bm{\eta}_{K}) \\
\end{array}
\right], \label{eqn:FIM}
\end{eqnarray}
where $\psi(\bm{\eta}_{k},\bm{\eta}_{k'})$ can be defined as,
\begin{eqnarray}
\psi(\bm{\eta}_{k},\bm{\eta}_{k'}) \triangleq -\mathbb{E}_{\mathbf{y},\bm{\eta}}\left[\frac{\partial^2 \ln p(\mathbf{y}|\boldsymbol{\bm{\eta}})}{\partial{\bm{\eta}_{k}}\partial{\bm{\eta}_{k'}}^T}\right]. \label{eqn:psi}
\end{eqnarray}

According to \cite{poor2013introduction}, the likelihood function of the received signal $\mathbf{y}$ conditioned on $\bm{\eta}$, $p(\mathbf{y}|\boldsymbol{\bm{\eta}})$, can be rewritten as,
\begin{eqnarray}
p(\mathbf{y}|\boldsymbol{\bm{\eta}}) \varpropto \exp \left\{ \frac{2}{N_0} \sum_{i=1}^{N}\mathbf{\mu}^{H}_{i}(\mathcal{L}, n)\mathbf{y}_{i}(\mathcal{L}, n) \right. \nonumber \\
\left. -\frac{1}{N_0}\sum_{i=1}^{N}\mathbf{\mu}^{H}_{i}(\mathcal{L}, n)^2 \right\},
\end{eqnarray}
where $\mathbf{\mu}^{H}_{i}(\mathcal{L}, n)=\sum_{k=0}^{K}\mathbf{h}_{i}^{(k)}(\mathcal{L}, n) \mathbf{x}_{i}(\mathcal{L}, n)$ and $\varpropto$ denotes equality up to irrelevant constants.

To obtain the CRLB of the proposed localization scheme, we transform the parameter vector $\bm{\eta}$ to $\tilde{\bm{\eta}}$, which includes the locations of scatters for $K$ fading paths, $\{\mathbf{s}_k\}$, and the modified parameter vector $\tilde{\bm{\eta}}$ can be obtained through,
\begin{eqnarray}
\mathbf{\tilde{\bm{\eta}}}=\left[\tilde{\bm{\eta}}_{0}^{T},\tilde{\bm{\eta}}_{1}^{T},\cdots\tilde{\bm{\eta}}_{k}^{T},\cdots \tilde{\bm{\eta}}_{K}^{T}\right]^{T},
\end{eqnarray}
where $\tilde{\bm{\eta}}_{k}=[\mathbf{s}_{k}^{T},\mathbf{h}_{k}^{T}]^{T}$ for $k>0$ and $\tilde{\bm{\eta}}_{0}=[\mathcal{L}_m^{T},\mathbf{h}_{0}^{T}]^{T}$ is the modified parameter vector for the direct path. Meanwhile, the FIM for $\tilde{\bm{\eta}}$, $\mathbf{J}_{\tilde{\bm{\eta}}}$, can be obtained by multiplying the transformation matrix $\mathbf{T}$, which gives
\begin{eqnarray}
\label{eqn:t3}
\mathbf{J}_{\tilde{\bm{\eta}}} = \mathbf{T}\mathbf{J}_{\bm{\eta}}\mathbf{T}^{T}.
\end{eqnarray}
Based on this transformation, the associated covariance error matrix for the target location $\mathcal{L}_m$ can be bounded as,
\begin{eqnarray}
\mathbb{E}_{\mathbf{y},\bm{\tilde{\eta}}}\left[ (\hat{\mathcal{L}_m}-\mathcal{L}_m)(\hat{\mathcal{L}_m}-\mathcal{L}_m)^{T}\right]\geq \left[\mathbf{J}_{\bm{\tilde{\eta}}}^{-1}\right]_{2\times2},
\end{eqnarray}
where $\left[\cdot\right]_{2\times2}$ denotes the projection operation to the $2\times2$ upper left sub-matrix. As a result, the CRLB of localization error using the proposed scheme can be obtained through the following theorem.
\begin{Thm}[CRLB of Localization Error]
\label{thm:peb}
The CRLB of localization error, $\epsilon_{\bm{\tilde{\eta}}}$, using the proposed MDE minimization framework, is given by,
\begin{eqnarray}
\epsilon_{\bm{\tilde{\eta}}} = \sqrt{tr\left\{\left[\mathbf{J}_{\bm{\tilde{\eta}}}^{-1}\right]_{2\times2}\right\}} = \sqrt{tr\left\{\left[\left(\mathbf{T}\mathbf{J}_{\bm{\eta}}\mathbf{T}^{T}\right)^{-1}\right]_{2\times2}\right\}},
\end{eqnarray}
where $\mathbf{J}_{\bm{\eta}}$ can be calculated through \eqref{eqn:FIM} and \eqref{eqn:psi}, and $\mathbf{T} \triangleq \partial\bm{\eta}^{T} / \partial\tilde{\bm{\eta}}^{T}$ as defined in \eqref{eqn:t1} and \eqref{eqn:t2}.
\end{Thm}

\begin{IEEEproof}[Proof of Theorem~\ref{thm:peb}]
Please refer to Appendix~\ref{appendix:matrix} for the proof.
\end{IEEEproof}

\subsection{Data Augmentation with Perturbation}
As mentioned before, a straightforward approach to improve the localization accuracy is to increase the number of $N_{RP}$, which generally requires careful measurement and labeling procedures in the offline training stage. In order to control the deployment complexity, we propose to use a perturbation based data augmentation scheme, where the perturbation distance $\|\Delta \mathcal{L}\|_2$ is much smaller than the localization distance $\|\hat{\mathcal{L}}_{m}\|_2$, i.e., $\|\Delta \mathcal{L}\|_2 \ll \|\hat{\mathcal{L}}_{m}\|_2$. Based on this perturbation, we can re-derive the CRLB of localization error as summarized below.

\begin{prop}[CRLB with Perturbation]
\label{prop:peb}
The CRLB of localization error with perturbation is denoted as,
\begin{eqnarray}
\epsilon_{\bm{\tilde{\eta}}_p} = \sqrt{tr\left\{\left[\mathbf{J}_{\bm{\tilde{\eta}},p}^{-1}\right]_{2\times2}\right\}} \approx \sqrt{tr\left\{\left[\left(\mathbf{J}_{\tilde{\bm{\eta}}} + \mathbf{J}_{\Delta\mathcal{L}}\right)^{-1}\right]_{2\times2}\right\}}.
\end{eqnarray}
where $\mathbf{J}_{\bm{\eta}}$ can be calculated through \eqref{eqn:t3}, and $\mathbf{J}_{\Delta\mathcal{L}}$ is defined in \eqref{eqn:c1}.
\end{prop}

\begin{IEEEproof}[Proof of Proposition~\ref{prop:peb}]
Please refer to Appendix~\ref{appendix:perturbation} for the proof.
\end{IEEEproof}

Since $tr\left\{\left[\mathbf{J}_{\Delta\mathcal{L},p}^{-1}\right]_{2\times2}\right\}\leq 0$ holds, then we have $\epsilon_{\bm{\tilde{\eta}}_p} \leq \epsilon_{\bm{\tilde{\eta}}}$, that is the operation of perturbation makes lower CRLB. Therefore the MDE minimization optimization problem is rewritten as follows.
\begin{Prob}[Logistic Regression with Augmentation]
\begin{eqnarray}
\underset{g (\cdot)}{\textrm{minimize}} &&\frac{1}{M} \sum_{m=1}^{M} \|\hat{\mathcal{L}}_m - (\mathcal{L}_m + \Delta\mathcal{L})\|_2 \\
 = && \frac{1}{M} \sum_{m=1}^{M} \left(\|\hat{\mathcal{L}}_m - \mathcal{L}_m\|_2+\alpha \cdot \|\Delta\mathcal{L}\|_2\right) \\
\textrm{subject to}
&&\hat{\mathcal{L}}_{m} \approx g \left(\left\{\mathbf{H}(\mathcal{L}_{m}+\Delta \mathcal{L},n_{m})\right\}\right), \\
&& \hat{\mathcal{L}}_{m}, \mathcal{L}_{m} \in \overline{\mathcal{L}_{RP}}, \forall m,
\end{eqnarray}
where $\alpha \in [-1, 1]$ denotes a fine-tuning coefficient and will be determined in the training stage.
\end{Prob}

With the above formulation, we numerically evaluate the CRLB to see how the position perturbation affect the CRLB and the corresponding numerical results are given in Section~\ref{sect:DA}.

\section{Deep Learning based Solution} \label{sect:sys}
In this section, we consider adopting classification and logistic regression based approaches to minimize the MDE mentioned in Section \ref{sect:prob} and design the neural network architecture for each of them. However, the corresponding difficulties are obvious. Firstly, under the situation of fluctuated wireless environments, the original sampled channel states contain unknown random noises, which may significantly degrade the estimation accuracy provided by neural networks. Secondly, the design methodology for logistic regression based localization is still unclear according to the existing literature. Last but not least, the deep learning based scheme usually requires huge amount of data to train the neural network parameters, which may incur significant overhead in the practical deployment. To address the above three challenges, we will introduce the proposed localization scheme in detail in this section.

\subsection{Data Collection and Cleaning}
\begin{figure}
\centering
\includegraphics[width = 3.3 in]{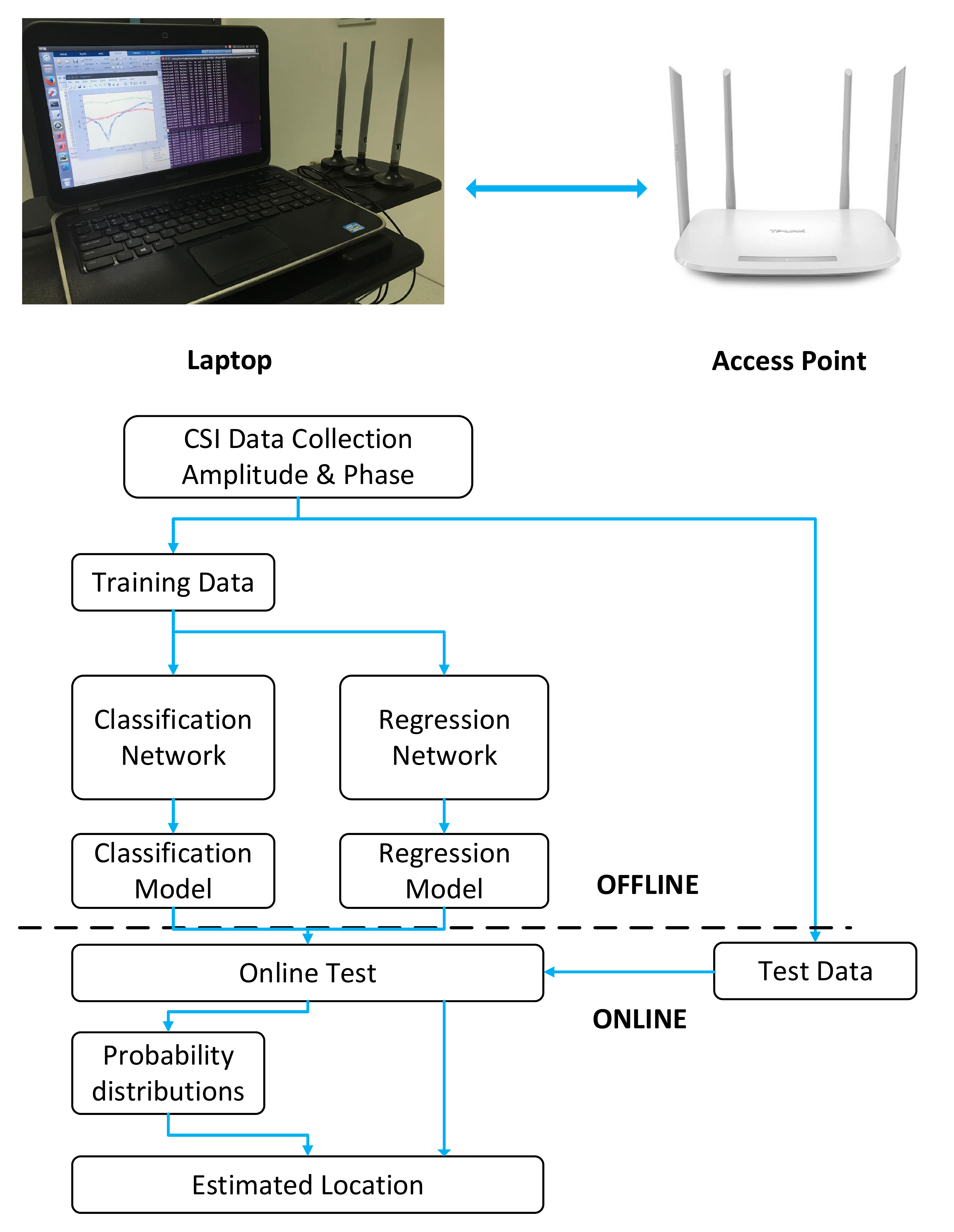}
\caption{The System Architecture, the whole process can be divided into offline phase and online phase, the amplitude and phase of collected CSI data are used as training and test data.}
\label{fig:system}
\end{figure}

Network Interface Cards (NICs) like Qualcomm Atheros AR series and Intel 5300 Nics make it possible to collect CSI data. Linux 802.11n CSI Tool \cite{halperin2010predictable} is the most widely used among the major CSI measurement tools. Consider a WiFi localization system as illustrated in Fig.~\ref{fig:system}, where the localization entity is a laptop equipped with Intel 5300 network interface card (NIC) and multiple receive antennas. The localization entity is working based on the real time receiving WiFi signals from an off-the-shelf AP with multiple transmit antennas.
Rather than generate data from numerical simulations, we collect the CSI data under 5.32GHz WiFi channel through a commercial laptop with multiple receiving antennas. CSI knowledge is computed on a WiFi OFDM symbol basis with duration 3.2$\mu$s according to IEEE 802.11n standards. The training dataset contains 60000\footnote{To accelerate the model training, we install Keras on our server with Intel(R) Xeon(R) CPU E5-3680 and NVIDIA Tesla P100 GPU.} transmitted packets and the packet interval is 4 ms, that is various channel situations of 4 minutes duration in the experimental environment are logged in the training dataset. Furthermore, CSI data extracted by the Linux 802.11n CSI Tool is transformed into polar coordinates for convenient data processing, i.e. $\mathbf{h}_{i}(\mathcal{L},n) = |\mathbf{h}_{i}(\mathcal{L},n)|e^{j\theta_{i}(\mathcal{L},n)}$, where $\left|\mathbf{h}_{i}(\mathcal{L}, n)\right|$ and $\theta_{i}(\mathcal{L},n)$ denote the corresponding amplitude and phase information respectively and $j$ represents the imaginary unit.

In the practical systems, the measured phase information, e.g. $\hat{\theta}_{i}(\mathcal{L},n)$ for subcarrier $i$ cannot be directly used for high accurate localization due to random jitters and noises caused by imperfect hardware components. In order to eliminate this effect, we adopt the common phase calibration algorithm proposed in \cite{xiao2012fifs}, and then obtain,
\begin{eqnarray}
\theta_{i}(\mathcal{L},n) = \hat{\theta}_{i}(\mathcal{L},n) + \frac{ 2\pi i}{N_{FFT}}\delta - Z,
\end{eqnarray}
where $N_{FFT}$ denotes the size of Fast Fourier Transform (FFT) \footnote{Linux 802.11n CSI Tool is designed according to IEEE 802.11n protocol, and the FFT size is 64.}, $\delta$ means the time lag at the receiver side, and $Z$ is unknown random measurement noise.

\subsection{Neural Network for Classification}
As mentioned in Problem~\ref{prob:CBL}, the localization accuracy rely on the accuracy of $\mathbf{p}_m (n_m)$ in such a classification problem. Common classification based deep learning neural network structures like MLP and CNN, are designed in this part to obtain the best approximation of $\mathbf{p}_m (n_m)$ and improve the final localization results.

Softmax function \cite{sutton2018reinforcement} is chosen as activation function at the output layer, which maps the output tensor values $\mathbf{a}=\left[a_1,a_2,\cdots,a_i,\cdots,a_{N_{RP}}\right]$ into the normalized prediction possibility $\mathbf{p}_m (n_m)$ in interval (0,1). The process can be described as,
\begin{eqnarray}
\mathbf{p}_{m,i}(n_m) =
\frac{e^{a_i}}{\sum_{i=1}^{N_{RP}}e^{a_i}},
\end{eqnarray}
where $\mathbf{p}_{m,i}$ is the $i^{th}$ element of the vector $\mathbf{p}_m (n_m)$. We also utilize {\em cross-entropy} as the loss function to measure the difference between the output normalized prediction $\mathbf{p}_{m,i}(n_m)$ and the true label vector $\mathbf{l}_{m,i}(n_m)$, which has proven to be a valid loss function for classification neural network \cite{de2005tutorial}. It can be written as,
\begin{eqnarray}
\mathbb{L} = -\sum_{i=1}^{N_{RP}}\mathbf{l}_{m,i}(n_m)\log\mathbf{p}_{m,i}(n_m),
\end{eqnarray}
where $\mathbf{l}_{m,i}(n_m)$ is the true label data for the $i^{th}$ RP location. Additionally, we train the parameters of deep neural networks with Stochastic Gradient Descent (SGD) method \cite{cherry2011saccharomyces} to minimize the loss function. In the online test phase, a probabilistic method utilize the estimated $\mathbf{p}_m (n_m)$ to obtain the final estimation location $\hat{\mathcal{L}}_m$ as mentioned in Problem \ref{prob:CBL}.

\subsection{Neural Network for Regression}
Our target is to find a better approximation of the non-convex function $g^{\star}_{LR}(\cdot)$ by the logistic regression based approach, so we choose the aggregated channel information, $\mathbf{H}(\mathcal{L}) = \big[\mathbf{H}(\mathcal{L}, 1),\ldots,$ $\mathbf{H}(\mathcal{L}, N)] \in \mathbb{C}^{N\times N_R \times N_{sc}}$, and the localization results, $\hat{\mathcal{L}}_{m} \in \mathbb{R}^{1\times2}$, to be the input and output matrices/vectors of neural networks respectively, and carefully select the loss function $\mathbb{L}$ to be the original definition of $g^{\star}_{LR}(\cdot)$,which is given by\footnote{To simplify the system implementation, we choose the number of sampling locations, $M$, to be equal to the number of RPs, $N_{RP}$.},
\begin{eqnarray}
\mathbb{L} = \frac{1}{N_{RP}} \sum_{m=1}^{N_{RP}} \|\hat{\mathcal{L}}_m - \mathcal{L}_m\|^2_2.
\end{eqnarray}
Kindly note that this type of loss function is quite different from {\em cross entropy} \cite{de2005tutorial} mentioned in classification based approach, which is often applied to describe the difference between the classification test results and the distribution of ground truth results. By minimizing the MDE with respect to the RPs, the logistic regression based approach can gradually converge to the non-convex function $g^{\star}_{LR}(\cdot)$ with satisfied performance via machine learning.

\begin{figure}
\centering
\includegraphics[width = 3.3 in]{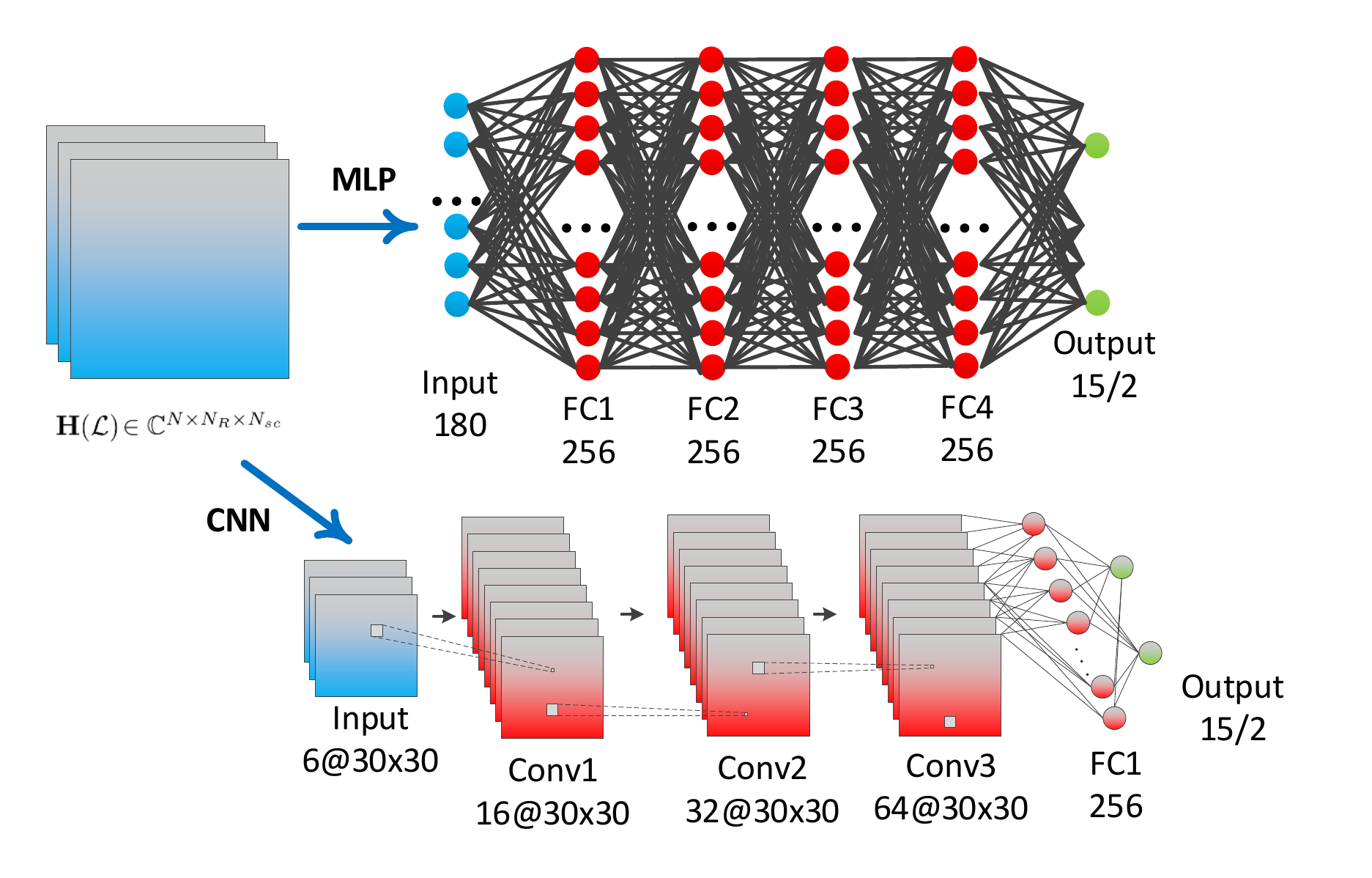}
\caption{The common architecture of MLP and CNN designed for classification and regression based neural network. MLP is consisted of fully-connected (FC) layers and CNN also contains convolution layers, pooling layers besides that.}
\label{fig:network}
\end{figure}

Additionally, to improve the identification and representation capability of neural networks, we exploit two classical neural networks with deeper structures, e.g., Multi-Layer Perception (MLP) and Convolutional Neural Networks (CNN), as shown in Fig.~\ref{fig:network}. Compared with the fully connection structure of MLP networks, the convolution layers in CNN provide the feasibility to extract the features from CSIs across the time and frequency domains, which may be more suitable for wireless fading environments as demonstrated in Section~\ref{sect:experiment}. To avoid the biasing effects caused by unusual samples, we apply the max pooling technique to deal with unimportant features and adopt the dropout technique \cite{srivastava2014dropout} to further reduce the unimportant connections in the neural networks. Meanwhile, to avoid the gradient vanishing problem, we also apply rectified linear unit (ReLU) \cite{nair2010rectified} as the non-linear activation functions in each hidden layers. The detailed configuration and parameters for both classification and regression networks are listed in Table~\ref{tab:parameter}.

\begin{table*} [ht]
\centering
\caption{An Overview of Network Configuration and Parameters.}
\label{tab:parameter}
\footnotesize
\begin{tabular}{c c c c}
\toprule
\textbf{Layers}&\textbf{MLP (Classification)}&\textbf{MLP (Regression)}&\textbf{CNN (Regression)}\\
\midrule
Input Layer & $180$ & $180$ & $6\times 30 \times 30$\\
\midrule
Hidden Layer 1  & FC 256 + ReLU & FC 256 + ReLU & Conv $16\times3\times3$ + ReLU\\
\midrule
Hidden Layer 2 & FC 256 + ReLU & FC 256 + ReLU& Conv $16\times3\times3$ + ReLU\\
\midrule
Hidden Layer 3  & FC 256 + ReLU & FC 256 + ReLU & \tabincell{c}{Conv $16\times3\times3$ + ReLU + MaxPooling}\\
\midrule
Hidden Layer 4 & \tabincell{c}{FC 64 + ReLU + Dropout 0.3}  & \tabincell{c}{FC 256 + ReLU + Dropout 0.3}& \tabincell{c}{FC 64 + ReLU + Dropout 0.3}\\
\midrule
Output Layer & FC 15 + Softmax & FC 2 + Linear & FC 2 + Linear\\
\midrule
Total No. of Para. & 257,574 & 248,322 & 236,114\\
\bottomrule
\end{tabular}
\end{table*}

\subsection{Outliers Removal}
Due to measurement uncertainty of online test signals, the matching process usually leads to a geographically dispersed set of test results $\{\hat{\mathcal{L}}_{m}\}$, resulting in unsatisfactory localization accuracy. Hence, the outlier points of the observations may exist in the set $\{\hat{\mathcal{L}}_{m}\}$, which is distant from other observations in statistical sense. In this paper, we proposed a outliers removal scheme to rule out the outlier points, which are far away from clustering set center in the decision process. We denote $(\bar x, \bar y)$ as the average point of set $\{\hat{\mathcal{L}}_{m}\}$, and $\textrm{std}_x$, $\textrm{std}_y$ as the standard deviation of $x$, $y$, which is defined as,
\begin{eqnarray}
\textrm{std}_x = \sqrt{\frac{1}{N_{gs}}\sum_{k=1}^{N_{gs}}(x_k-\bar x)^2}, \\
\textrm{std}_y = \sqrt{\frac{1}{N_{gs}}\sum_{k=1}^{N_{gs}}(y_k-\bar y)^2},
\end{eqnarray}
where $N_{gs}$ is the sample group size. If the following conditions,
\begin{eqnarray}
\delta_x=\frac{|x_k-\bar x|}{\textrm{std}_x}>\delta_{th}, \\
\delta_y=\frac{|y_k-\bar y|}{\textrm{std}_y}>\delta_{th},
\end{eqnarray}
are held, then $(x_k,y_k)$ is considered as a outlier point and removed from the set $\{\hat{\mathcal{L}}_{m}\}$, in which $\delta_{th} > 0$ is the designed rejection threshold.

\section{Experiment Results} \label{sect:experiment}
\begin{figure}[t]
\centering
\subfigure[Laboratory Scenario]{
\includegraphics[width=2.3 in]{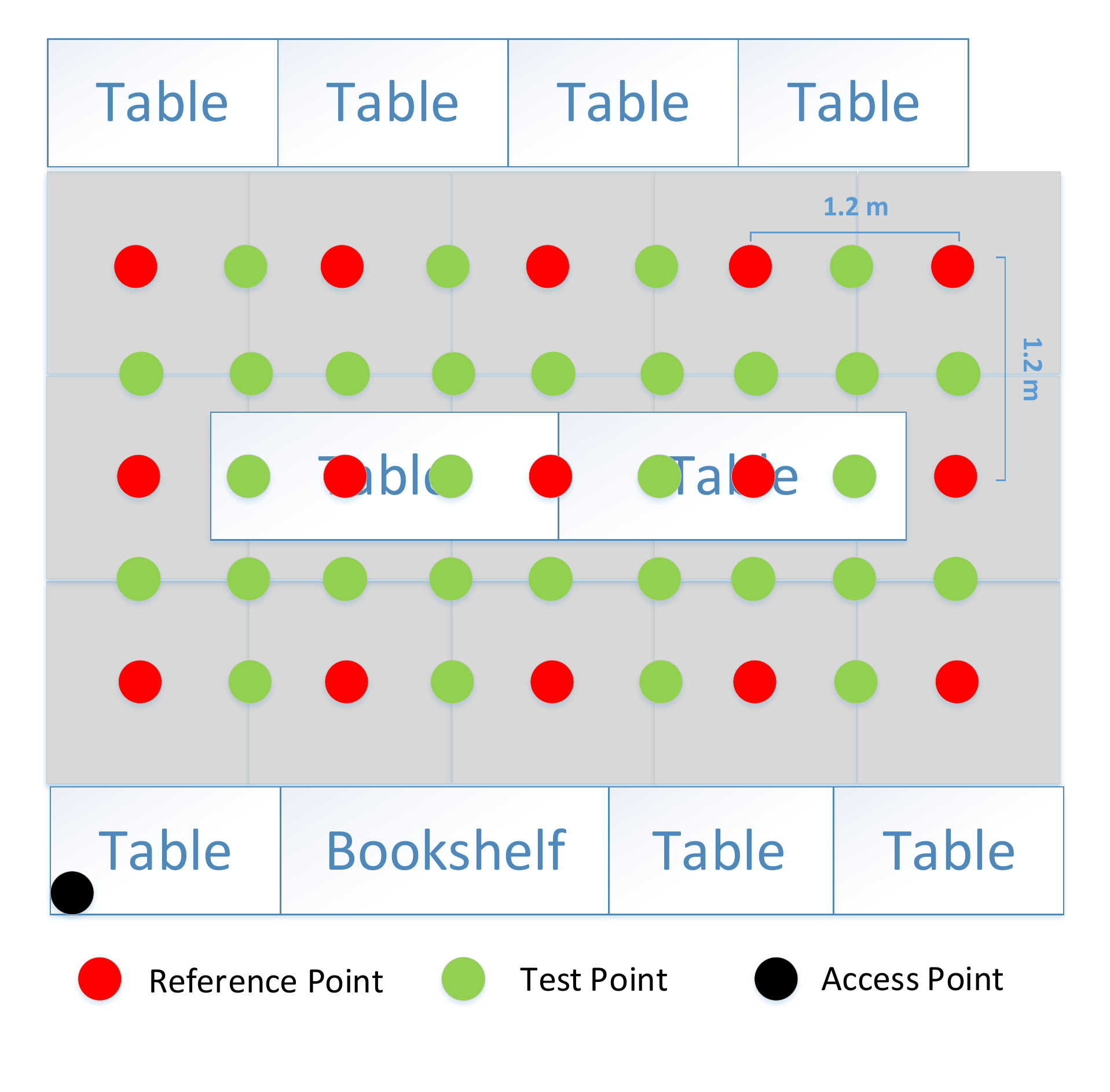}}
\hfill
\centering
\subfigure[Corridor Scenario]{
\includegraphics[width=3.4 in]{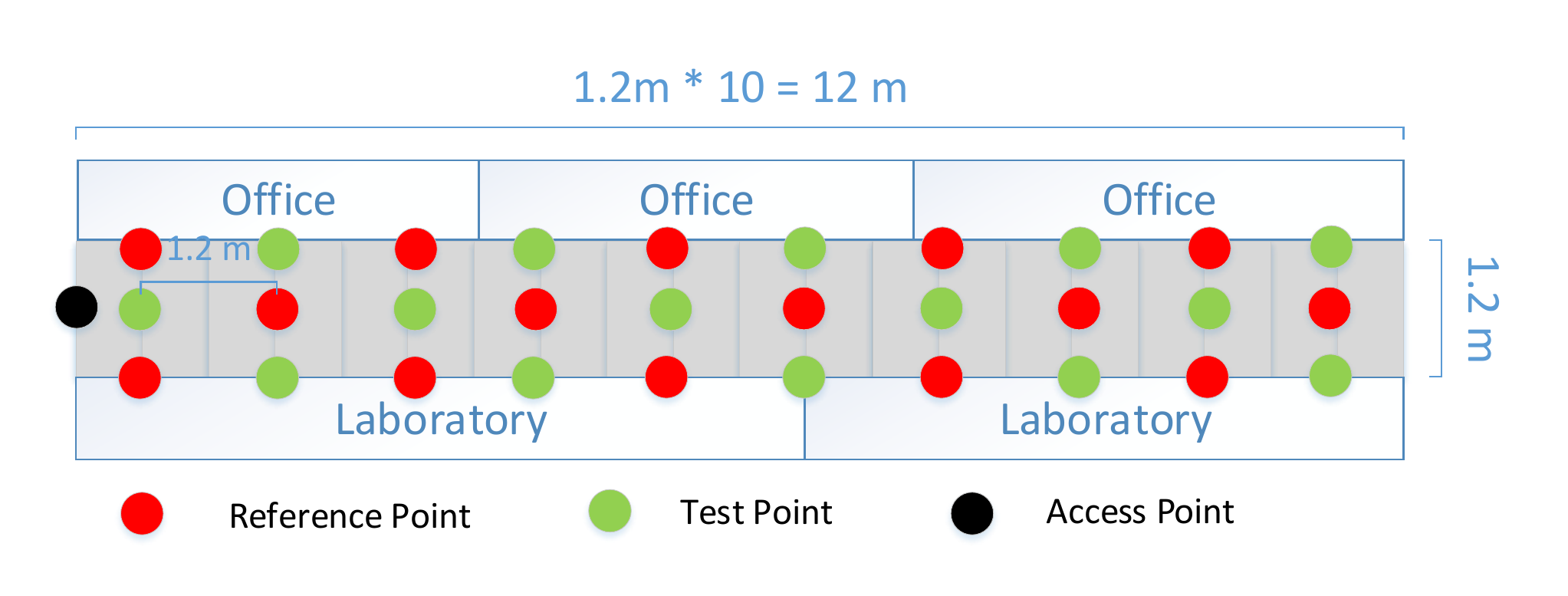}}
\caption{A sketch map of experiment environment, including laboratory scenario and corridor scenario, where the red, green, black spots represent the location of RPs, test points (TPs) and APs respectively. The distance between two adjacent RPs is 1.2m.}
\label{fig:environment}
\end{figure}

In this section, we provide some numerical results to show the effectiveness of the proposed logistic regression based approach for indoor localization. More specifically, we compare the proposed scheme with two baseline systems, e.g., {\em Baseline 1}: KNN based localization scheme and {\em Baseline 2}: classification based localization scheme with MLP architecture. We verify the proposed logistic regression based localization scheme in both laboratory and corridor environment, where the layout of testing scenarios are shown in Fig.~\ref{fig:environment}. With laboratory equipment, furniture, and people movements in the real situation, the tested wireless fading conditions cover most of the daily indoor scenarios with mixed LOS and NLOS paths.

\subsection{Logistic Regression vs. Classification}
\begin{figure}[t]
\centering
\subfigure[Laboratory Scenario]{
\includegraphics[width=3.4 in]{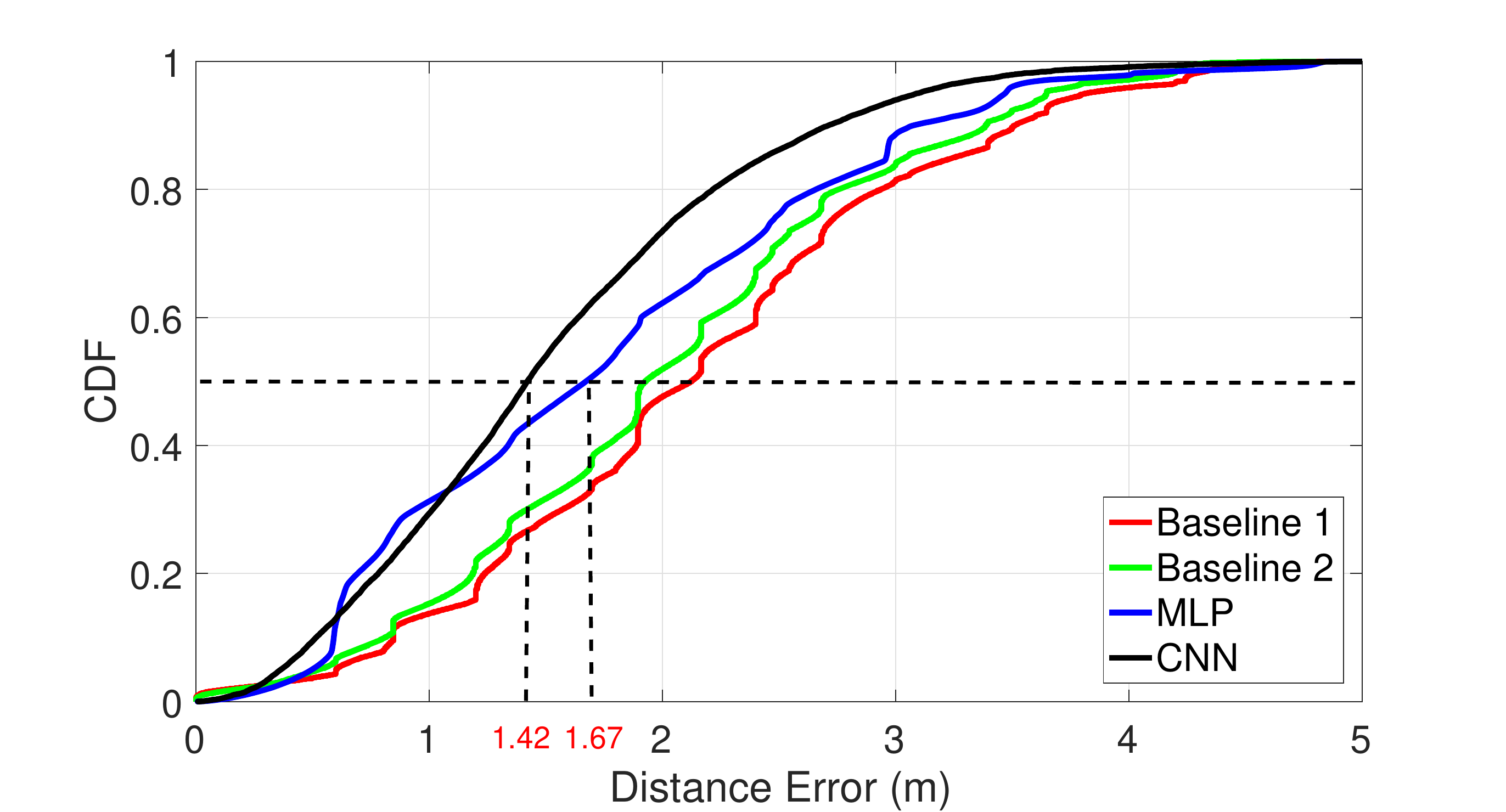}}
\hfill
\centering
\subfigure[Corridor Scenario]{
\includegraphics[width=3.4 in]{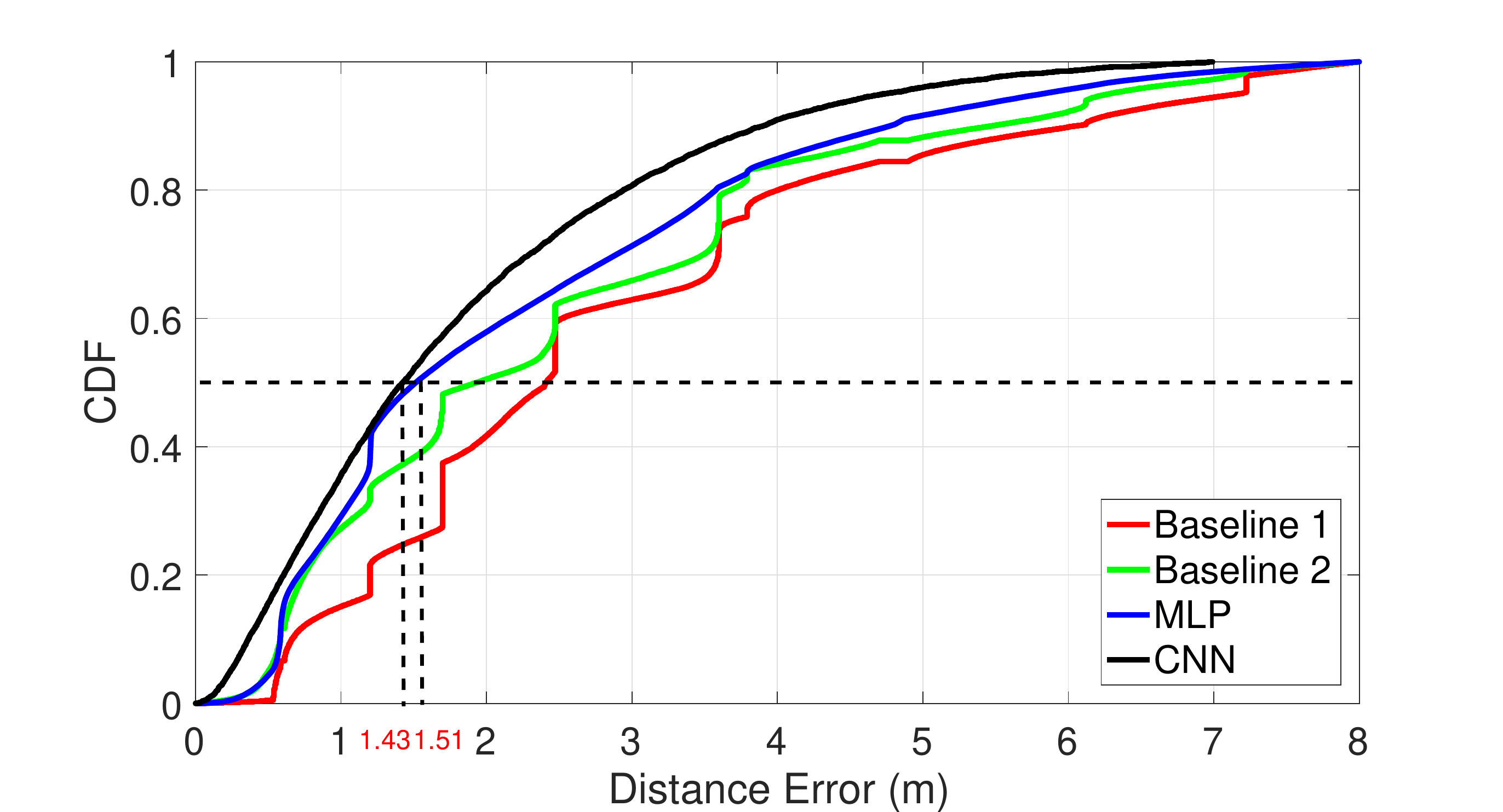}}
\caption{CDF of localization distance errors for different algorithms in both laboratory and corridor scenarios. The proposed regression based approach with MLP and CNN architectures are compared with two baselines to test the algorithm effectiveness.}
\label{fig:Distance error}
\end{figure}

In the first experiment, we compared the proposed regression scheme with the above baselines by measuring the cumulative distribution function (CDF) of distance error in the laboratory scenario, as well as the corridor scenario. Fig.~\ref{fig:Distance error} describes CDF of the localization distance error during the operating stage. The proposed regression based algorithms show superior localization accuracy over conventional algorithms, including KNN based localization ({\em Baseline 1}) and classification based localization ({\em Baseline 2}) for both two cases.

By comparing MLP-based approach (blue solid curves) and CNN-based approach (black solid curves), the latter one achieves the median errors of 1.42m and 1.43m for the laboratory and corridor scenarios respectively, which shows better localization accuracy than the former one (1.67m in laboratory case and 1.51m in corridor case). This is due to the fact that CNN-based approach is able to capture the time domain correlations of multiple OFDM symbols, while MLP-based approach only focuses on extracting the common features among all the observations. Kindly note that due to the similar numbers of parameters, the time complexity for Baseline 2, MLP and CNN based methods is similar as well\footnote{KNN based method (Baseline 1) will cost much more time than others, due to the absence of the training process.}, which is 0.11s, 0.11s, 0.12s for each test.

\subsection{Effect of System Configuration}
In this experiment, we investigate on the effect of system configurations, like the APs number and grid size, which will directly affect deployment cost and localization accuracy. We would like to find a practical tradeoff between deployment cost and localization accuracy, and explore the most efficient deployment setting for our system.

\begin{figure}
\centering
\includegraphics[width = 3.4 in]{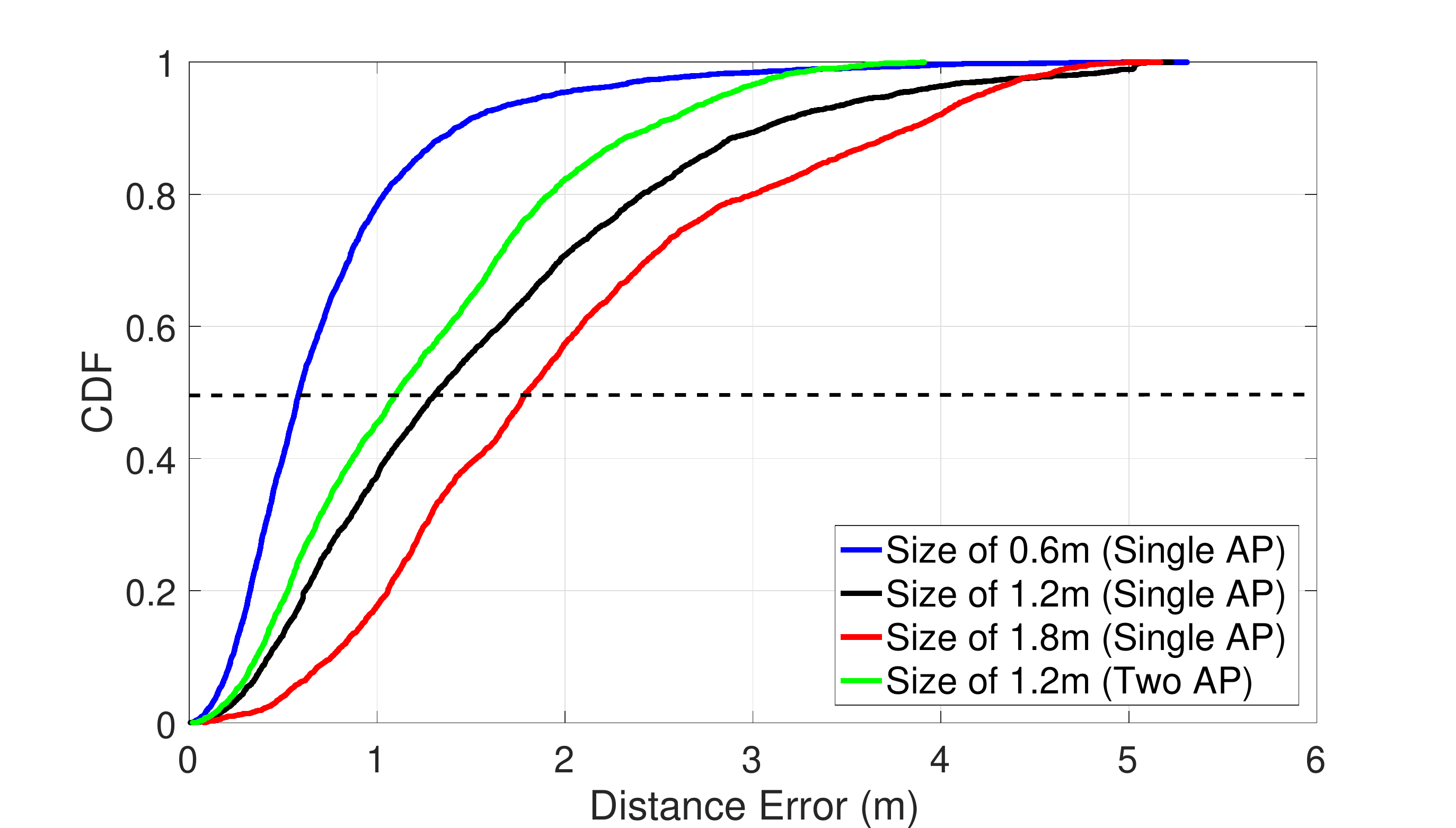}
\caption{CDF of localization errors for different AP number and grid sizes to in the laboratory scenario using regression based approach with CNN architecture.}
\label{fig:config}
\end{figure}

In the above experiments, the distance between the adjacent RPs each is set as 1.2m and only one WiFi AP is deployed. Hence, we add a new AP placed at a different corner of the laboratory and add different grid size of 0.6m and 1.8m as supplementary. The location error results of different system configurations are illustrated in Fig.~\ref{fig:config}.
We find that the average distance error of each APs are 1.42m (green curves) and 1.51m respectively, and up to 1.10m when using both of them\footnote{Two independent APs receive transmitted signals and log the CSI data at the same time. The output location result of each AP will be simply averaged to verify the performance of two APs setting.}, improving 23$\%$ compared with using single AP. Furthermore, using two APs can effectively reduce the possibility of large location error more than 3m, which greatly improve the system reliability. We leave the question that how to better use the relationship between two APs for the future work, rather than simply taking the average results from multiple APs.

The median location accuracy of different grid size with 1.8m, 1.2m and 0.6m settings are 1.86m, 1.42m and 0.61m respectively. It is worth noting that when the size is changed from 1.2m to 0.6m, the localization accuracy improve 57\% at the cost of four times of data collection and labeling. We conclude that the median distance errors are close to the corresponding grid size. It means that the operators can choose the deployment grid size according to the accuracy that he wants to achieve, which has a certain guiding significance for the actual deployment.

\subsection{Effect of Outliers Removal}
\begin{figure}
\centering
\includegraphics[width = 3.4 in]{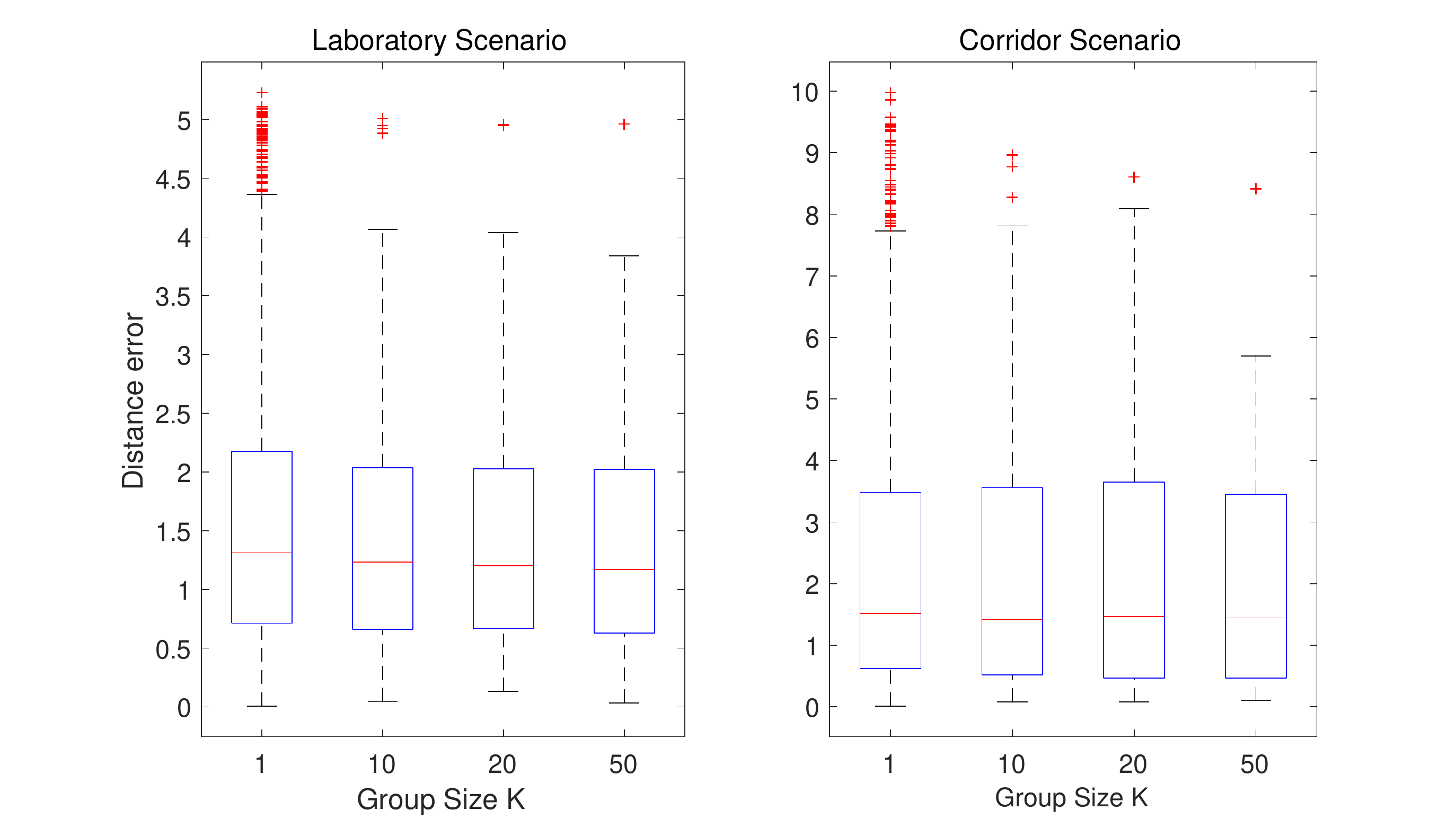}
\caption{Boxplot of localization errors in laboratory scenario and corridor scenario with different group sizes after outliers removal algorithm. }
\label{fig:group size}
\end{figure}

The above CDF plots are drawn according to the results of each online test, without any post processing techniques. In this part, the proposed outliers removal scheme is used to rule out the abnormal test results and bring out much better localization accuracy. Two boxplots\footnote{In each box, the red line indicates the median, and the bottom and top edges of the box indicate the $25^{th}$ and $75^{th}$ percentiles respectively. The whiskers extend to the most extreme data, and the outliers are plotted individually using the `+' symbol.} of location error in the laboratory scenario and corridor scenario are shown in Fig.~\ref{fig:group size}. We make $N_{gs}$ equal to 10, 20, 50 as the group size of the test results in our system and rule out the abnormal coordinate results in each group. Eventually, the average results of each group are calculated to represent this group.

As shown in Fig.~\ref{fig:group size}, we find that the maximum error of our localization system has been controlled within 4m if 10 continuous results taken as a group. Furthermore, if 20 or 50 continuous results are taken as a group, the maximum error is only about 1m, which greatly increases the reliability of our system. The location errors have been effectively reduced after using our outliers removal algorithm. The time cost of every fusion test are 1.27s, 2.54s, 6.35s for group size of 10, 20 and 50. Considering the time cost versus location accuracy, 20 is the most reasonable group size scheme among these three settings for our system. Therefore we can conclude that taking advantage of outlier removal techniques will bring better robustness and reliability of localization system.

\subsection{Effect of Data Augmentation} \label{sect:DA}
\begin{figure}[t]
\centering
\includegraphics[width = 3.3 in]{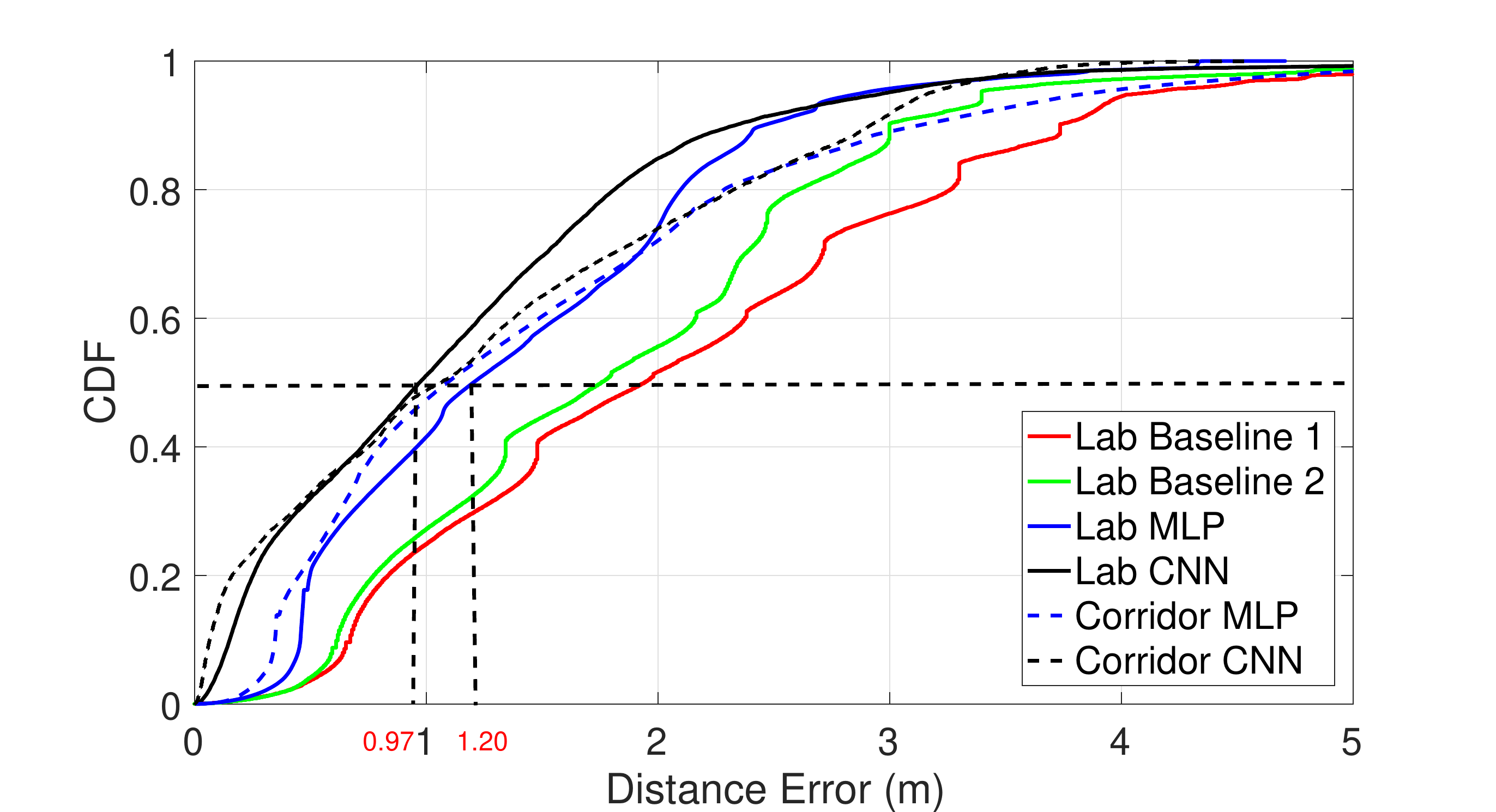}
\caption{CDF of localization errors for four different algorithms, including {\em Baseline 1}, {\em Baseline 2} and the proposed logistic regression with MLP and CNN architecture. Newly collected CSI datasets in laboratory and corridor environment are used to verify the effectiveness of data augmentation.}
\label{fig:augmentation}
\end{figure}

In order to verify the effectiveness of data augmentation with perturbation, we extend the original training dataset by adding some perturbed samples with $\Delta \mathcal{L}$ less than 0.1m. We re-train the neural networks with the augmented dataset and redo the same experiments in the operating stage. The CDF of distance errors for different algorithms are illustrated in Fig.~\ref{fig:augmentation}.

Under the effect of data augmentation, the positioning accuracy of both MLP-based approach (blue solid curves) and CNN-based approach (black solid curves) have been obviously improved, e.g., from 1.67m to 1.20m and from 1.42m to 0.97m for the median distance error in laboratory scenario, which corresponds to 28\% and 32\% improvement respectively. In the corridor scenario, the proposed data augmentation scheme shows the similar improvement on localization accuracy as well, which verifies the effectiveness of data augmentation with perturbation as mentioned in Section~\ref{sect:prob}.

\section{Conclusion} \label{sect:conc}
In this paper, we propose a logistic regression based localization scheme for WiFi systems. By applying a unified optimization framework, we compare it with the conventional classification based approaches, and derive the corresponding CRLB of the localization errors. Based on the analytical results, we find that using perturbations to extend the training dataset can improve the CRLB accordingly. Together with some outlier removal techniques, we show through numerical experiments that the proposed logistic regression based approach is shown to be effective for the localization accuracy improvement, which achieves robust sub-meter level MDE using a single WiFi AP.

\begin{appendices}
\numberwithin{equation}{section}
\makeatletter
\newcommand{\section@cntformat}{Appendix \thesection:\ }
\makeatother
\section{Proof of Hypothesis 1}
\label{appendix:proof_hyp}
To obtain the only determinable position, we borrow the idea of successive interference cancellation (SIC) technique \cite{patel1994analysis} to estimate the signal component of each path step by step from the received time domain signal $\mathbf{y}(t)$. For example, we estimate the parameters $\bm{\eta}_{0}$ from the original received signal $\mathbf{y}(t)$. After the first estimation, we reconstruct signal $\mathbf{s}_0(t)$ using $\bm{\eta}_{0}$,
\begin{eqnarray}
\mathbf{s}_{0}(t) = h_{0}\cdot\mathbf{A}_{R}(t) \mathbf{A}^{\textrm{H}}_{T}(t) U(t-\tau),
\end{eqnarray}
where $U(t)\in \mathbb{C}^{N_T}$ is the step function. After cancelling $\mathbf{s}_{0}(t)$ from $\mathbf{y}(t)$, the residual signal $\mathbf{y}_{1}(t)$ is calculate as,
\begin{eqnarray}
\mathbf{y}_{1}(t) &=& \mathbf{y}(t) - \mathbf{s}_{0}(t) \nonumber \\
&=& \sum_{k=1}^K\mathbf{s}_{k}(t)+ \mathbf{n}(t),
\end{eqnarray}
Similarly, the second strongest signal $\mathbf{s}_{1}(t)$ can be obtained from the $\mathbf{y}_{1}(t)$ residual signal. We iterate this process until signals of each paths are separately estimated, stopping when the power of the residual signal $\mathbf{y}_{K}(t)$ is below the radio noise range. The parameter $\bm{\eta}_{0}$ estimated from the received signal of the LOS path $\mathbf{s}_{0}(t)$, including AOA $\theta_{R,0}$ and the straight distance $d_0 = c \cdot \tau_{0}$, as well as the AP location $\mathcal{L}_{0}=(x_0, y_0)$, help to estimate $\mathcal{L}_{m}$, which is expressed as,
\begin{eqnarray}
\mathcal{L}_{m} = \left(x_{0}+d_{0}\cos\theta_{R,0}, y_{0}+d_{0}\sin\theta_{R,0}\right)^{T}.
\end{eqnarray}
If the multi-path components are taken into consideration, $\mathcal{L}_{m}$ can be rewritten as,
\begin{eqnarray}
\mathcal{L}_{m} = \frac{1}{K+1}\sum_{k=0}^{K+1}(x_{0}+d_{k,1}\cos\theta_{T,k}-d_{k,2}\cos\theta_{R,k}, \nonumber \\
y_{0}+d_{k,1}\sin\theta_{T,k}+d_{k,2}\sin\theta_{R,k})^{T}.
\end{eqnarray}

\section{Proof of Theorem 1}
\label{appendix:matrix}
We obtain the entries of $\mathbf{T}$ from the geometry relations between the parameters $\eta$ and $\tilde\eta$ as illustrated in Fig.~\ref{fig:multieffect}, which is given as follows.
\begin{eqnarray}
\tau_{0}&=&\frac{1}{c}\|\mathcal{L}_m-\mathcal{L}_{0}\|_{2}, \nonumber \\
\tau_{k}&=&\|\mathcal{L}_{0}-\mathcal{L}^{s}_{k}\|_{2} /c+\|\mathcal{L}_m-\mathcal{L}^{s}_{k}\|_{2} /c, \nonumber \\
\theta_{T,0}&=&\arccos((x_m-x_0)/\|\mathcal{L}_m-\mathcal{L}_{0}\|_{2}), \nonumber \\
\theta_{T,k}&=&\arccos((x_s-x_0)/\|\mathcal{L}_m-\mathcal{L}_{0}\|_{2}), \nonumber \\
\theta_{R,k}&=&\pi-\arccos((x_m-x_s)/\|\mathcal{L}_m-\mathcal{L}^{s}_{k}\|_{2}), \nonumber \\
\theta_{R,0}&=&\pi+\arccos((x_m-x_0)/\|\mathcal{L}_m-\mathcal{L}_{0}\|_{2}),
\end{eqnarray}
where $k>0$. Consequently, we obtain,
\begin{eqnarray}
\mathbf{T} = \left[
\begin{array}{c c c}
\mathbf{T}_{0,0}&\cdots &\mathbf{T}_{K,0}\\
 \vdots  & \ddots  & \vdots  \\
\mathbf{T}_{0,K}&\cdots &\mathbf{T}_{K,K} \\
\end{array}
\right],
\end{eqnarray}
where $\mathbf{T}_{k,k'}$ is defined as,
\begin{eqnarray}
\mathbf{T}_{k,k'} \triangleq \frac{\partial\bm{\eta}_{k}^{T}}{\partial\tilde{\bm{\eta}_{k'}}}.
\end{eqnarray}
For $k'\neq 0$, $\mathbf{T}_{k,k'}$ is obtained as,
\begin{eqnarray}
\mathbf{T} = \left[
\begin{array}{c c c}
\partial\bm{\tau}_{k}^{T} /\partial \mathcal{L}_{m}
&\partial\bm{\theta}_{k}^{T} /\partial \mathcal{L}_{m} &\partial\mathbf{h}_{k}^{T} /\partial \mathcal{L}_{m}\\
\partial\bm{\tau}_{k}^{T} /\partial \mathbf{h}_{k'}
&\partial\bm{\theta}_{k}^{T} /\partial \mathbf{h}_{k'} &\partial\mathbf{h}_{k}^{T} /\partial \mathbf{h}_{k'}\\
\end{array}
\right], \label{eqn:t1}
\end{eqnarray}
and $\mathbf{T}_{k,0}$ is obtained as
\begin{eqnarray}
\mathbf{T} = \left[
\begin{array}{c c c}
\partial\bm{\tau}_{k}^{T} /\partial \mathcal{L}^{s}_{k'}
&\partial\bm{\theta}_{k}^{T} /\partial \mathcal{L}^{s}_{k'} &\partial\mathbf{h}_{k}^{T} /\partial \mathcal{L}^{s}_{k'}\\
\partial\bm{\tau}_{k}^{T} /\partial \mathbf{h}_{0}
&\partial\bm{\theta}_{k}^{T} /\partial \mathbf{h}_{0} &\partial\mathbf{h}_{k}^{T} /\partial \mathbf{h}_{0}\\
\end{array}
\right], \label{eqn:t2}
\end{eqnarray}
where
\begin{eqnarray}
\partial\bm{\tau}_{0}/\partial \mathcal{L}_{m} &=& \frac{1}{c}[\cos\theta_{T,0},\sin\theta_{T,0}]^T, \nonumber \\
\partial\theta_{T,0}/\partial \mathcal{L}_{m} &=& \frac{1}{\|\mathcal{L}_m-\mathcal{L}_{0}\|_{2}}[-\sin\theta_{T,0},\cos\theta_{T,0}]^T, \nonumber \\
\partial\theta_{R,0}/\partial \mathcal{L}_{m} &=& \frac{1}{\|\mathcal{L}_m-\mathcal{L}_{0}\|_{2}}[-\sin\theta_{T,0},\cos\theta_{T,0}]^T, \nonumber\\
\partial\bm{\tau}_{k}/\partial \mathcal{L}_{m} &=& \frac{1}{c}[-\cos\theta_{R,k},-\sin\theta_{R,k}]^T,
\end{eqnarray}
and
\begin{eqnarray}
\partial\bm{\tau}_{k}/\partial \mathcal{L}^{s}_{k} = \frac{1}{c}[\cos\theta_{T,k}+\cos\theta_{R,k}, \sin\theta_{T,k}+\sin\theta_{R,k}]^T, \nonumber \\
\partial\theta_{T,k}/\partial \mathcal{L}^{s}_{k}= \frac{1}{\|\mathcal{L}^{s}_{k}-\mathcal{L}_{0}\|_{2}}[-\sin\theta_{T,k},\cos\theta_{T,k}]^T, \nonumber \\
\partial\theta_{R,k}/\partial \mathcal{L}^{s}_{k}= -\frac{1}{\|\mathcal{L}^{s}_{k}-\mathcal{L}_{m}\|_{2}}[\sin\theta_{T,k},-\cos\theta_{T,k}]^T,
\end{eqnarray}
for $k>0$, and the rest of entries in $\mathbf{T}$ are zero.

\section{Proof of Proposition 1}
\label{appendix:perturbation}
Under the assumption of $\|\Delta \mathcal{L}\|_2 \ll \|\hat{\mathcal{L}}_{m}\|_2$, we have
\begin{eqnarray}
\hat{\mathcal{L}}_{m} \approx \hat{\mathcal{L}}_{m} + \Delta \mathcal{L} = g \left(\left\{\mathbf{H}(\mathcal{L}_{m}+\Delta \mathcal{L},n_{m})\right\}\right),
\end{eqnarray}
that is we can collect more CSI samples without changing the location labels. Since the norm satisfies the triangle inequality, we have
\begin{eqnarray}
\|\hat{\mathcal{L}}_m - \mathcal{L}_m\|_2 - \|\Delta\mathcal{L}\|_2 \leq \|\hat{\mathcal{L}}_m - (\mathcal{L}_m + \Delta\mathcal{L})\|_2 \nonumber \\
\leq \|\hat{\mathcal{L}}_m - \mathcal{L}_m\|_2+ \|\Delta\mathcal{L}\|_2.
\end{eqnarray}

To explain the role of data augmentation, we re-derive the FIM and CRLB under the effect of the perturbation. We define $\mathcal{L}_p = \mathcal{L}_m + \Delta\mathcal{L}$ is the position under perturbation distance $\|\Delta\mathcal{L}\|_2$, and then
the unknown channel parameter under perturbation is defined as,
\begin{eqnarray}
\mathbf{\tilde{\bm{\eta}}}_p=\left[\tilde{\bm{\eta}}_{p,0}^{T},\tilde{\bm{\eta}}_{p,1}^{T},\cdots\tilde{\bm{\eta}}_{p,k}^{T},\cdots \tilde{\bm{\eta}}_{p,K}^{T}\right]^{T},
\end{eqnarray}
where $\tilde{\bm{\eta}}_{p,0}=[\mathcal{L}_m^{T},\mathbf{h}_{0}^{T},{\Delta\mathcal{L}}^{T}]^{T}$, and $\tilde{\bm{\eta}}_{p,k}=[\mathbf{s}_{k}^{T},\mathbf{h}_{k}^{T},{\Delta\mathcal{L}}^{T}]^{T}$ for $k>0$. Based on the above assumption of $\|\Delta \mathcal{L}\|_2 \ll \|\hat{\mathcal{L}}_{m}\|_2$, the FIM for $\mathbf{\tilde{\bm{\eta}}}_p$ with the presence of random parameters is given by \cite{van2004detection},
\begin{eqnarray}
\mathbf{J}_{\tilde{\bm{\eta}},p} \approx \mathbf{J}_{\tilde{\bm{\eta}}} + \mathbf{J}_{\Delta\mathcal{L}},
\end{eqnarray}
where $\mathbf{J}_{\Delta\mathcal{L}}$ can be calculated as,
\begin{eqnarray}
\label{eqn:c1}
\mathbf{J}_{\Delta\mathcal{L}} = -\mathbb{E}_{\mathbf{\tilde{\bm{\eta}}}_p}\left[\frac{\partial^2 \ln p(\mathbf{\tilde{\bm{\eta}}}_p)}{\partial\mathbf{\tilde{\bm{\eta}}}_p\partial\mathbf{\tilde{\bm{\eta}}}_p^{\textrm{T}}}\right],
\end{eqnarray}
and hence the CRLB of localization error with perturbation is denoted as,
\begin{eqnarray}
\epsilon_{\bm{\tilde{\eta}}_p} = \sqrt{tr\left\{\left[\mathbf{J}_{\bm{\tilde{\eta}},p}^{-1}\right]_{2\times2}\right\}}.
\end{eqnarray}
\end{appendices}

\bibliographystyle{IEEEtran}
\bibliography{IEEEabrv,wifi}

\begin{thebibliography}{10}
\providecommand{\url}[1]{#1}
\csname url@samestyle\endcsname
\providecommand{\newblock}{\relax}
\providecommand{\bibinfo}[2]{#2}
\providecommand{\BIBentrySTDinterwordspacing}{\spaceskip=0pt\relax}
\providecommand{\BIBentryALTinterwordstretchfactor}{4}
\providecommand{\BIBentryALTinterwordspacing}{\spaceskip=\fontdimen2\font plus
\BIBentryALTinterwordstretchfactor\fontdimen3\font minus
  \fontdimen4\font\relax}
\providecommand{\BIBforeignlanguage}[2]{{%
\expandafter\ifx\csname l@#1\endcsname\relax
\typeout{** WARNING: IEEEtran.bst: No hyphenation pattern has been}%
\typeout{** loaded for the language `#1'. Using the pattern for}%
\typeout{** the default language instead.}%
\else
\language=\csname l@#1\endcsname
\fi
#2}}
\providecommand{\BIBdecl}{\relax}
\BIBdecl

\bibitem{xiang2019robust}
C.~Xiang, Z.~Zhang, S.~Zhang, S.~Xu, S.~Cao, and V.~LAU, ``Robust sub-meter
  level indoor localization-a logistic regression approach,'' in \emph{IEEE
  Proc. ICC'19}, 2019, pp. 1--6.

\bibitem{chintalapudi2010indoor}
K.~Chintalapudi, A.~Padmanabha~Iyer, and V.~N. Padmanabhan, ``Indoor
  localization without the pain,'' in \emph{ACM Proc. MobiCom'10}, 2010, pp.
  173--184.

\bibitem{marais2005land}
J.~Marais, M.~Berbineau, and M.~Heddebaut, ``Land mobile gnss availability and
  multipath evaluation tool,'' \emph{{IEEE} Trans. Veh. Technol.}, vol.~54,
  no.~5, pp. 1697--1704, 2005.

\bibitem{ijaz2013indoor}
F.~Ijaz, H.~K. Yang, A.~W. Ahmad, and C.~Lee, ``Indoor positioning: A review of
  indoor ultrasonic positioning systems,'' in \emph{IEEE Proc. ICACT'13}, 2013,
  pp. 1146--1150.

\bibitem{altini2010bluetooth}
M.~Altini, D.~Brunelli, E.~Farella, and L.~Benini, ``Bluetooth indoor
  localization with multiple neural networks,'' in \emph{IEEE Proc. ISWPC'10},
  2010, pp. 295--300.

\bibitem{bouet2008rfid}
M.~Bouet and A.~L. Dos~Santos, ``{RFID} tags: Positioning principles and
  localization techniques,'' in \emph{IEEE Proc. IFIP'08}, 2008, pp. 1--5.

\bibitem{hauschildt2010advances}
D.~Hauschildt and N.~Kirchhof, ``Advances in thermal infrared localization:
  Challenges and solutions,'' in \emph{IEEE Proc. IPIN'10}, 2010, pp. 1--8.

\bibitem{youssef2005horus}
M.~Youssef and A.~Agrawala, ``The {H}orus {WLAN} location determination
  system,'' in \emph{ACM Proc. MobiSys'05}, 2005, pp. 205--218.

\bibitem{kotaru2015spotfi}
M.~Kotaru, K.~Joshi, D.~Bharadia, and S.~Katti, ``Spot{F}i: Decimeter level
  localization using {W}i{F}i,'' in \emph{ACM Proc. SIGCOMM'15}, vol.~45,
  no.~4, 2015, pp. 269--282.

\bibitem{vasisht2016decimeter}
D.~Vasisht, S.~Kumar, and D.~Katabi, ``Decimeter-level localization with a
  single {W}i{F}i {A}ccess {P}oint.'' in \emph{ACM Proc. NSDI'16}, vol.~16,
  2016, pp. 165--178.

\bibitem{sen2012you}
S.~Sen, B.~Radunovic, R.~R. Choudhury, and T.~Minka, ``You are facing the
  {M}ona {L}isa: spot localization using phy layer information,'' in \emph{ACM
  Proc. MobiSys'12}, 2012, pp. 183--196.

\bibitem{chapre2015csi}
Y.~Chapre, A.~Ignjatovic, A.~Seneviratne, and S.~Jha, ``C{SI}-{MIMO}: An
  efficient {W}i-{F}i fingerprinting using channel state information with
  {MIMO},'' \emph{Pervasive and Mobile Computing}, vol.~23, pp. 89--103, 2015.

\bibitem{wang2017biloc}
X.~Wang, L.~Gao, and S.~Mao, ``Biloc: Bi-modal deep learning for indoor
  localization with commodity 5{GH}z {W}i{F}i,'' \emph{IEEE Access}, vol.~5,
  pp. 4209--4220, 2017.

\bibitem{wang2015deepfi}
X.~Wang, L.~Gao, S.~Mao, and S.~Pandey, ``Deep{F}i: Deep learning for indoor
  fingerprinting using channel state information,'' in \emph{IEEE Proc.
  WCNC'15}, 2015, pp. 1666--1671.

\bibitem{chen2017confi}
H.~Chen, Y.~Zhang, W.~Li, X.~Tao, and P.~Zhang, ``Con{F}i: Convolutional neural
  networks based indoor {W}i-{F}i localization using channel state
  information,'' \emph{IEEE Access}, vol.~5, pp. 18\,066--18\,074, 2017.

\bibitem{wang2017resloc}
X.~Wang, X.~Wang, and S.~Mao, ``Resloc: Deep residual sharing learning for
  indoor localization with {CSI} tensors,'' in \emph{IEEE Proc. PIMRC'17},
  2017, pp. 1--6.

\bibitem{peduzzi1996simulation}
P.~Peduzzi, J.~Concato, E.~Kemper, T.~R. Holford, and A.~R. Feinstein, ``A
  simulation study of the number of events per variable in logistic regression
  analysis,'' \emph{Jour. Clin. Epide.}, vol.~49, no.~12, pp. 1373--1379, 1996.

\bibitem{kay1993fundamentals}
S.~M. Kay, \emph{Fundamentals of statistical signal processing}.\hskip 1em plus
  0.5em minus 0.4em\relax Prentice Hall PTR, 1993.

\bibitem{shahmansoori2017position}
A.~Shahmansoori, G.~E. Garcia, G.~Destino, G.~Seco-Granados, and H.~Wymeersch,
  ``Position and orientation estimation through millimeter-wave mimo in 5g
  systems,'' \emph{{IEEE} Trans. Wireless Commun.}, vol.~17, no.~3, pp.
  1822--1835, 2017.

\bibitem{bevington1993data}
P.~R. Bevington, D.~K. Robinson, J.~M. Blair, A.~J. Mallinckrodt, and S.~McKay,
  ``Data reduction and error analysis for the physical sciences,''
  \emph{Computers in Physics}, vol.~7, no.~4, pp. 415--416, 1993.

\bibitem{cai2017cril}
S.~Cai, W.~Liao, C.~Luo, M.~Li, X.~Huang, and P.~Li, ``{CRIL}: An efficient
  online adaptive indoor localization system,'' \emph{{IEEE} Trans. Veh.
  Technol.}, vol.~66, no.~5, pp. 4148--4160, 2017.

\bibitem{belagiannis2015robust}
V.~Belagiannis, C.~Rupprecht, G.~Carneiro, and N.~Navab, ``Robust optimization
  for deep regression,'' in \emph{IEEE Proc. ICCV'05}, 2015, pp. 2830--2838.

\bibitem{amemiya1985advanced}
T.~Amemiya, \emph{Advanced econometrics}.\hskip 1em plus 0.5em minus
  0.4em\relax Harvard university press, 1985.

\bibitem{frieden2004science}
B.~R. Frieden, \emph{Science from Fisher information: a unification}.\hskip 1em
  plus 0.5em minus 0.4em\relax Cambridge University Press, 2004.

\bibitem{poor2013introduction}
H.~V. Poor, \emph{An introduction to signal detection and estimation}.\hskip
  1em plus 0.5em minus 0.4em\relax Springer Science \& Business Media, 2013.

\bibitem{halperin2010predictable}
D.~Halperin, W.~Hu, A.~Sheth, and D.~Wetherall, ``Predictable 802.11 packet
  delivery from wireless channel measurements,'' in \emph{ACM Proc.
  SIGCOMM'10}, vol.~40, no.~4, 2010, pp. 159--170.

\bibitem{xiao2012fifs}
J.~Xiao, K.~Wu, Y.~Yi, and L.~M. Ni, ``{FIFS}: Fine-grained indoor
  fingerprinting system.'' in \emph{IEEE Proc. ICCCN'12}, 2012, pp. 1--7.

\bibitem{sutton2018reinforcement}
R.~S. Sutton and A.~G. Barto, \emph{Reinforcement learning: An
  introduction}.\hskip 1em plus 0.5em minus 0.4em\relax MIT press, 2018.

\bibitem{de2005tutorial}
P.-T. De~Boer, D.~P. Kroese, S.~Mannor, and R.~Y. Rubinstein, ``A tutorial on
  the cross-entropy method,'' \emph{Annals. Operation. Res}, vol. 134, no.~1,
  pp. 19--67, 2005.

\bibitem{cherry2011saccharomyces}
J.~M. Cherry, E.~L. Hong, C.~Amundsen, R.~Balakrishnan, G.~Binkley, E.~T. Chan,
  K.~R. Christie, M.~C. Costanzo, S.~S. Dwight, S.~R. Engel \emph{et~al.},
  ``Saccharomyces genome database: the genomics resource of budding yeast,''
  \emph{Nucleic acids research}, vol.~40, no.~D1, pp. D700--D705, 2011.

\bibitem{srivastava2014dropout}
N.~Srivastava, G.~Hinton, A.~Krizhevsky, I.~Sutskever, and R.~Salakhutdinov,
  ``Dropout: a simple way to prevent neural networks from overfitting,''
  \emph{Journ. Mach. Learn. Res.}, vol.~15, no.~1, pp. 1929--1958, 2014.

\bibitem{nair2010rectified}
V.~Nair and G.~E. Hinton, ``Rectified linear units improve restricted boltzmann
  machines,'' in \emph{Proc. ICML'10}, 2010, pp. 807--814.

\bibitem{patel1994analysis}
P.~Patel and J.~Holtzman, ``Analysis of a simple successive interference
  cancellation scheme in a ds/cdma system,'' \emph{{IEEE} J. Sel. Areas
  Commun.}, vol.~12, no.~5, pp. 796--807, 1994.

\bibitem{van2004detection}
H.~L. Van~Trees, \emph{Detection, estimation, and modulation theory, part I:
  detection, estimation, and linear modulation theory}.\hskip 1em plus 0.5em
  minus 0.4em\relax John Wiley \& Sons, 2004.

\end{thebibliography}
\EOD
\end{document}